\documentclass[
  journal=pasa,
  manuscript=research-paper, 
  year=2022,
  volume=37,
]{cup-journal}

\usepackage{microtype,siunitx,booktabs}

\usepackage{hyperref}

\newcommand{\barolo}{\textsc{3DBarolo}}

\newcommand{\fat}{\textsc{FAT}}

\newcommand{\tirific}{\textsc{TiRiFiC}}

\newcommand{\sofia}{\textsc{SoFiA}}

\newcommand{\mcgsuite}{\textsc{MCGSuite}}

\newcommand{\wkapp}{\textsc{WKAPP}}

\newcommand{\hi}{\text{H\sc{i}}}

\newcommand{\ellmaj}{\code{ell\_maj}}

\newcommand{\arcsec}{$''$}

\newcommand{\kms}{\textrm{km s}^{-1}}

\newcommand{\code}[1]{\texttt{#1}}

\defcitealias{SoFiARelease}{W22}

\newcommand{\ICRAR}[0]{International Centre for Radio Astronomy Research (ICRAR), The University of Western Australia, 35 Stirling Highway, Crawley WA 6009, Australia}
\newcommand{\ASTRO}[0]{ARC Centre of Excellence for All Sky Astrophysics in 3~Dimensions (ASTRO~3D), Australia}
\newcommand{\CSIROPERTH}[0]{CSIRO Space and Astronomy, PO Box 1130, Bentley WA 6102, Australia}
\newcommand{\CSIROSYDNEY}[0]{ATNF, CSIRO Space and Astronomy, PO Box~76, Epping NSW~1710, Australia}
\newcommand{\AUSSRC}[0]{Australian SKA Regional Centre (AusSRC)}

\title{WALLABY Pilot Survey: Public release of HI kinematic models for more than 100~galaxies from phase~1 of ASKAP pilot observations}

\author{N.~Deg}
\affiliation{Department of Physics, Engineering Physics, and Astronomy, Queen's University, Kingston ON K7L~3N6, Canada}

\author{K. Spekkens}
\affiliation{Department of Physics and Space Science, Royal Military College of Canada, P.O.\ Box 17000, Station Forces Kingston ON K7K~7B4, Canada}

\author{T. Westmeier}
\affiliation{\ICRAR}
\alsoaffiliation{\ASTRO}

\author{T.N. Reynolds}
\affiliation{\ICRAR}
\alsoaffiliation{\ASTRO}

\author{P. Venkataraman}
\affiliation{Dunlap Institute of Astronomy and Astrophysics, University of Toronto, 50 St. George Street, Toronto, ON, M5S 3H4, Canada}

\author{S. Goliath}
\affiliation{NRC Herzberg Astronomy and Astrophysics Research Centre, 5071 W. Saanich Rd., Victoria, BC, V9E 2E7, Canada}

\author{A.~X.~Shen}
\affiliation{\CSIROPERTH}
\alsoaffiliation{\AUSSRC}

\author{R. Halloran}
\affiliation{Department of Physics, Engineering Physics, and Astronomy, Queen's University, Kingston ON K7L~3N6, Canada}

\author{A. Bosma}
\affiliation{Aix Marseille Univ, CNRS, CNES, LAM, Marseille}

\author{B.  Catinella}
\affiliation{\ICRAR}
\alsoaffiliation{\ASTRO}

\author{W.J.G. de Blok}
\affiliation{Netherlands Institute for Radio Astronomy (ASTRON), Oude Hoogeveensedijk 4, 7991 PD Dwingeloo, The Netherlands}
\alsoaffiliation{Department of Astronomy, University of Cape Town, Private Bag X3, Rondebosch 7701, South Africa}
\alsoaffiliation{Kapteyn Astronomical Institute, University of Groningen, PO Box 800, 9700 AV Groningen, The Netherlands}

\author{H.~D\'{e}nes}
\affiliation{Netherlands Institute for Radio Astronomy (ASTRON), Oude Hoogeveensedijk~4, 7991 PD Dwingeloo, The Netherlands}

\author{E. M. Di Teodoro}
\affiliation{Department of Physics \& Astronomy, Johns Hopkins University, Baltimore, MD 21218, USA}
\alsoaffiliation{Space Telescope Science Institute, 3700 San Martin Drive, Baltimore, MD 21218, USA}

\author{A.~Elagali}
\affiliation{Telethon Kids Institute, Perth Children's Hospital, Perth, Australia}

\author{B.-Q.~For}
\affiliation{\ICRAR}
\alsoaffiliation{\ASTRO}

\author{C.~Howlett}
\affiliation{School of Mathematics and Physics, The University of Queensland, Brisbane QLD~4072, Australia}

\author{G.~I.~G. J{\'o}zsa}
\affiliation{Max-Planck-Institut f\"ur Radioastronomie, Auf dem H\"ugel 69, D-53121 Bonn, Germany}
\alsoaffiliation{Department of Physics and Electronics, Rhodes University, P.O. Box 94, Makhanda, 6140, South Africa}

\author{P.~Kamphuis}
\affiliation{Ruhr University Bochum, Faculty of Physics and Astronomy, Astronomical Institute, 44780~Bochum, Germany}

\author{D.~Kleiner}
\affiliation{INAF -- Osservatorio Astronomico di Cagliari, Via della Scienza~5, 09047~Selargius, CA, Italy}

\author{B. Koribalski}
\affiliation{\CSIROSYDNEY}
\alsoaffiliation{School of Science, Western Sydney University, Locked Bag~1797, Penrith NSW~2751, Australia}

\author{K.~Lee-Waddell}
\affiliation{\ICRAR}
\alsoaffiliation{\CSIROPERTH}

\author{F. Lelli}
\affiliation{INAF - Arcetri Astrophysical Observatory, Largo Enrico Fermi 5, 50125, Florence Italy}

\author{X. Lin}
\affiliation{School of Physics, Peking University, Beijing 100871, People's Republic of China}

\author{C.~Murugeshan}
\affiliation{\CSIROPERTH}
\alsoaffiliation{\ASTRO}

\author{S. Oh}
\affiliation{Department of Astronomy and Space Science, Sejong University, 209, Neungdong-ro, Gwangjin-gu, Seoul, Republic of Korea}
\alsoaffiliation{Department of Physics and Astronomy, Sejong University, 209, Neungdong-ro, Gwangjin-gu, Seoul, Republic of Korea}

\author{J.~Rhee}
\affiliation{\ICRAR}
\alsoaffiliation{\ASTRO}

\author{T. C. Scott}
\affiliation{Instituto de Astrofísica e Ci\^{e}ncias do Espa\c{c}o (IA), Rua das Estrelas, 4150-762 Porto, Portugal}

\author{L.~Staveley-Smith}
\affiliation{\ICRAR}
\alsoaffiliation{\ASTRO}

\author{J.M. van der Hulst}
\affiliation{Kapteyn Astronomical Institute, University of Groningen, PO Box 800, 9700 AV Groningen, The Netherlands}

\author{L. Verdes-Montenegro}
\affiliation{Instituto de Astrofísica de Andalucía (IAA-CSIC), Glorieta de la Astronomía, 18008 Granada, Spain}

\author{J.~Wang}
\affiliation{Kavli Institute for Astronomy and Astrophysics, Peking University, Beijing~100871, China}

\author{O.~I.~Wong}
\affiliation{\CSIROPERTH}
\alsoaffiliation{\ICRAR}
\alsoaffiliation{\ASTRO}

\doi{article in press}

\received {23 June 2022}
\revised  {24 Aug 2022}
\accepted {02 Sept 2022}
\published{: currently in press}


\begin{document}

\begin{abstract}
We present \textcolor{black}{the} Widefield ASKAP L-band Legacy All-sky Blind surveY (WALLABY) Pilot Phase I \hi\ kinematic models.  \textcolor{black}{
This first data release consists of \hi\ observations of three fields in the direction of the Hydra and Norma clusters, and the NGC 4636 galaxy group.  In this paper, we describe how we generate and publicly  release flat-disk tilted-ring kinematic models for 109/592 unique \hi\ detections in these fields.  The modelling method adopted here -- which we call the WALLABY Kinematic Analysis Proto-Pipeline (WKAPP) and for which the corresponding scripts are also publicly available -- consists of combining results from the homogeneous application of the \fat\ and \barolo\ algorithms \textcolor{black}{to the subset of 209 detections with sufficient resolution and $S/N$ in order} to generate optimized model parameters and uncertainties. 
The 109 models presented here tend to be gas rich detections resolved by at least 3--4 synthesized beams across their major axes, but there is no obvious environmental bias in the modelling. The data release described here is the first step towards the derivation of similar products for thousands of spatially-resolved WALLABY detections via a dedicated kinematic pipeline.   Such a large publicly available and homogeneously analyzed dataset will be a powerful legacy product that that will enable a wide range of scientific studies.  }
\end{abstract}

\section{Introduction}
\label{sec:int}

The Widefield ASKAP L-band Legacy All-sky Blind surveY (WALLABY; \citealt{Koribalski2020}) is one of the key science projects for the Australian SKA Pathfinder (ASKAP; \citealt{Hotan2021}) telescope.  It is an extragalactic survey expected to detect the atomic hydrogen (\hi) gas content of $\sim 210,000$ galaxies out to redshift $z \sim 0.1$.  It is expected that thousands of these sources will have sufficient spatial resolution for kinematic modelling.  WALLABY has completed the first phase of its pilot observations, consisting of the three fields towards the Hydra and Norma clusters, and the NGC 4636 group.  \citet[][hereafter W22]{SoFiARelease}  is the release paper for the pilot data release 1 (PDR1) and this paper describes the PDR1 \textcolor{black}{rotating disk} kinematic models.

The generation of reliable kinematic models for as many resolved galaxies as possible is a key science driver for WALLABY.  \textcolor{black}{Most, but not all, sources with significant \hi\ reservoirs are rotationally supported as the gas generally settles into rotating disks due to the conservation of angular momentum.  As such, we attempt to model all the sufficiently resolved PDR1 detections using `rotating disk' kinematic models}.
Such models provide important measurements for galaxies that are useful for exploring a variety of questions.  For instance, the rotation curves generated from such models are key to answering questions related to the mass distribution \textcolor{black}{within} galaxies.  Such questions include whether or not disks are maximal \citep{vanAlbada1985,vanAlbada1986,Lelli2016,Starkman2018}, and, with a large enough sample size, probing the core-cusp problem \citep{deBlok2010}.  Additionally, studies of the Tully-Fisher (TF) relation \citep{Tully1977} are significantly improved by measurements of $v_{\rm{flat}}$ (which is derived from the outer rotation curves; see \citealt{Verheijen2001,Ponomareva2016,Lelli2019}).  Getting a statistically significant sample of $v_{\rm{flat}}$ measurements will be valuable for galaxy population studies involving the TF relation and the baryonic TF relation \citep{McGaugh2000,Oh2011}.

Another key use of these rotation models is the calculation of the resolved velocity function down to low masses \citep{Lewis2019}.  This, coupled with mass modelling will help to address questions related to the dark matter (DM) halo-to-\hi\ velocity relation \citep{Papastergis2016}.  For larger galaxies, kinematic models can help to constrain their spin, warps, and angular momentum.  The DM spin and \hi\ warps of a galaxy are often connected to environmental processes \citep{Battaner1990,Stevens2016,Lagos2018}.

These are just a few of the questions that require robust \hi\ kinematic models.  There are a variety of different methods for generating such kinematic models from interferometric observations. The most common is tilted-ring modelling \citep{Rogstad1974}, initially applied in 2D to velocity moments of well-resolved \hi\ detections  \citep[e.g.][]{bosma78, Begeman1987,vanderHulst1992}. More recently, algorithms have been developed to apply tilted-ring models directly to 3D datacubes \citep[e.g.][]{Jorza2007,Davis2013,Kamphuis2015,diTeodoro15,Bekiaris2016}. A key advantage of 3D techniques relative to their 2D counterparts is the reliability with which they can be applied to marginally spatially-resolved \hi\ detections; thus making them particularly useful for homogeneous application to large numbers of detections from blind widefield surveys such as WALLABY.

This paper describes \textcolor{black}{the construction of the PDR1 kinematic models (for the subset of galaxies that were successfully modelled) as well as the public release of the resulting data products.}
In Sec. \ref{sec:obs} we briefly describe the WALLABY detections, but a full description is provided in the data release paper \citepalias{SoFiARelease}.  Sec.~\ref{sec:TRModelling} describes tilted ring modelling in general, while \textcolor{black}{Sec.}~\ref{Sec:ModelProcedure} describes the specific approach taken for the PDR1 observations.  Sec.~\ref{Sec:CatalogueAndProducts} provides the overall results of the PDR1 kinematic models.  \textcolor{black}{Sec.}~\ref{Sec:Populations} describes the population of kinematically modelled galaxies, and Sec.~\ref{Sec:Conclusions} provides the conclusions and discusses the future of kinematic modelling for full WALLABY.

\section{WALLABY PDR1 Detections}
\label{sec:obs}

The WALLABY pilot phase 1 observations targeted three $60\,\mathrm{deg}^2$ fields that cover cluster and group environments at differing distances $D$: the Hydra cluster ($D \sim 60\,$Mpc, \citealt{Jorgensen1996, Reynolds2021}), the Norma cluster ($D \sim 70\,$Mpc, \citealt{Mutabazi2021}), and the NGC 4636 group ($D \sim 15\,$Mpc, \citealt{tully2013}). 
A full description of the observations, the data reduction and the application of the \textsc{SoFiA} source finding code \textcolor{black}{(the HI Source Finding Application; }\citealt{Serra2015, Westmeier2021}) to generate the PDR1 sample of 592 unique \hi\ detections divided between Hydra Team Release 1 (TR1), Hydra TR2, Norma TR1 and NGC~4636 TR1 is reported in \citetalias{SoFiARelease}.  

 
 The PDR1 detection cubelets \textcolor{black}{(cutouts from the mosaiced cubes around each detected source)}, imaged with a Gaussian restoring beam with a full-width at half maximum of $30''$ (hereafter ``the beam") in $18.5\,$kHz-wide spectral channels ( $=3.9\,\kms$ at $z=0$), are the starting point for the kinematic analysis presented here.  \citetalias{SoFiARelease} also detail the limitations of the  data given the pilot nature of the observations; we discuss \textcolor{black}{the} potential impacts of the those limitations on the kinematic models in Section~\ref{subsec:mod_lim}, which we expect to be mild.



\begin{figure}[t]
\centering
    \includegraphics[width=\textwidth]{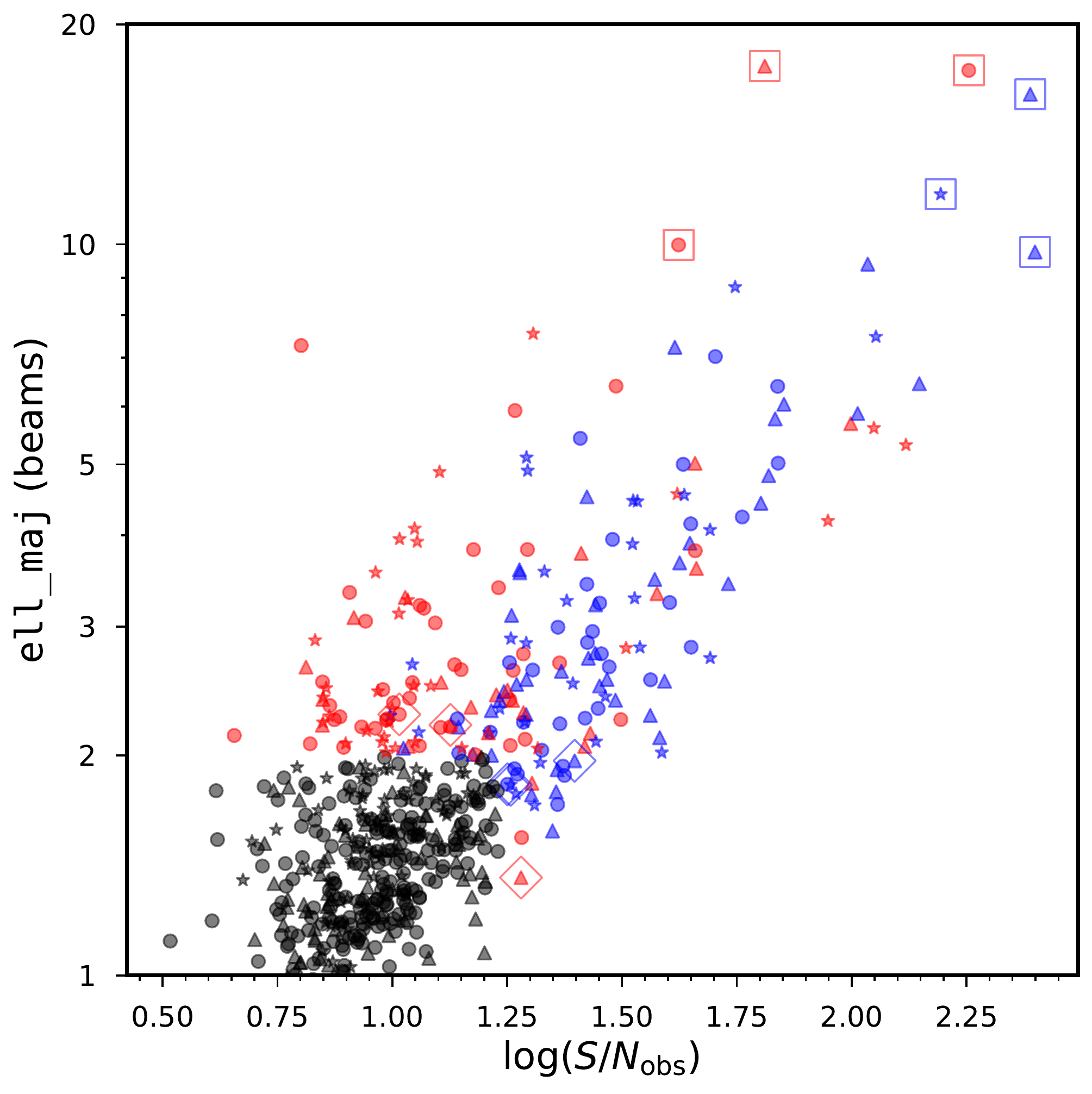} \caption{Size (as estimated by \code{ell\_maj}) as a function of integrated $S/N$ (given by Eq.~\ref{Eq:SN}) of PDR1 detections. Sources in the Hydra, Norma, and NGC~4636 fields are indicated by circles, stars, and triangles respectively. \textcolor{black}{Coloured points represent all detections with $\code{ell\_maj} > 2\,$beams or $\log(S/N_{obs}) > 1.25$, for which kinematic models were attempted: successful models are shown in blue, and failed models are shown in red (see Section~\ref{Sec:ModelProcedure})}.  Moment maps for the sources corresponding to the points outlined in larger open symbols are shown in Fig.~\ref{Fig:Montage}.}
  \label{Fig:Size-SNPlot}
\end{figure}

Figure~\ref{Fig:Size-SNPlot} plots the angular size as a function of the integrated signal-to-noise ($S/N$) for the detected sources.  These two properties strongly influence whether a galaxy's \hi\ content can be reliably kinematically modelled (see Section~\ref{Sec:CatalogueAndProducts}). We define the size in Fig.~\ref{Fig:Size-SNPlot} as the SoFiA-returned major axis diameter \code{ell\_maj} of an ellipse fitted to the source Moment 0 map \citep{Westmeier2021}. As in \citetalias{SoFiARelease}, we compute the integrated $S/N$ via:
\begin{equation}\label{Eq:SN}
    S/N_{obs}=\frac{S_{mask}}{\sigma_{rms} \sqrt{N_{mask} \Omega}}~,
\end{equation}
where $S_{mask}$ and $N_{mask}$ are the total flux and number of cells in the SoFiA detection mask respectively, $\Omega$ is the beam area, and $\sigma_{rms}$ is the root-mean-square noise of the detection-free cells in the corners of the SoFiA-returned source cubelet. In Fig.~\ref{Fig:Size-SNPlot} and throughout, we plot the Hydra TR2 values for sources with both Hydra TR1 and Hydra TR2 entries.

Fig.~\ref{Fig:Size-SNPlot} shows a clear correlation between angular size and $S/N$ among the detections: as expected, sources with larger angular sizes have higher integrated $S/N$. As also shown in \citetalias{SoFiARelease}, the majority of the detections are only marginally spatially resolved, with values of \code{ell\_maj} that span only a few beams.  Moreover, most detections have relatively low $S/N$. As such, our modelling  must be tailored for the marginally resolved, low $S/N$ regime. \textcolor{black}{We discuss considerations that drive the adopted modelling approach in Sec.~\ref{sec:TRModelling}, and describe the resulting procedure in Sec.~\ref{Sec:ModelProcedure}.}

\section{Kinematic Modelling Considerations}
\label{sec:TRModelling}

Given that many science goals for WALLABY are enabled by statistical samples of resolved source properties (see Sec.~\ref{sec:int}),  two core principles underpin our kinematic modelling approach: 
\begin{enumerate}
\item Models should be automatically and homogeneously applied to all suitable detections. 
\item Model parameters should have robust estimates of their uncertainties.
\end{enumerate}
These principles drive key choices in the modelling undertaken. First, we do not tailor kinematic models to individual detections; rather, we apply the same models using the same technique to all sources that meet our selection criteria. Second, since available algorithms do not return statistical uncertainties on all parameters, we apply different code implementations of the same underlying model to a given source in order to estimate the uncertainties for the returned parameters.

Given these principles and the properties of the spatially-resolved PDR1 detections described in Sec.~\ref{sec:obs}, we discuss here the considerations that drive our kinematic modelling procedure. Sec.~\ref{subsec:TRmodcodes} introduces tilted-ring modelling and describes the \textcolor{black}{Fully Automated \tirific\ (FAT, where \tirific\ itself stands for Tilted Ring Fitting Code; \citealt{Kamphuis2015, Jorza2007})} and \textcolor{black}{the 3D-Based Analysis of Rotating Objects From Line Observations (\barolo, \citealt{diTeodoro15})} algorithms that we use to generate the PDR1 models. 
Sec.~\ref{Sec:ModelCodeComp} then explores  differences in how these two codes model the same underlying observation,  which is used to build and hone the modelling procedure adopted in Sec.~\ref{Sec:ModelProcedure}.


\subsection{Tilted-Ring Modelling}
\label{subsec:TRmodcodes}

Tilted-ring modelling, first introduced by \citet{Rogstad1974}, is a widely-used technique for generating kinematic models of a galaxy's \hi\ disk.  In this procedure, a model galaxy is constructed from a series of concentric rings, each with intrinsic properties such as a centre, rotation speed, surface density and thickness, as well as quantities that arise from the ring's sky projection, like inclination and position angle.  While the precise set of parameters included in the models varies by implementation, the goal is to generate mock observations of the ring ensemble and optimize the ring parameters so that they resemble the observations.  

Tilted-ring models were initially developed for application to 2D velocity fields derived from 3D \hi\ datacubes, with \textsc{rotcur} in the \textsc{gipsy} package being an early and widely-used implementation  \citep{Begeman1987,vanderHulst1992}. A suite of more recent 2D algorithms that also characterize non-circular flows or complex disk geometries have since been developed and publicly released (e.g.\ \textsc{reswri}, \citealt{Schoenmakers97}; \textsc{Kinemetry}, \citealt{Krajnovic2005};  \textsc{DiskFit}, \citealt{Spekkens2007}; \textsc{2DBAT}; \citealt{Oh2018}). 2D algorithms are relatively efficient, and reliably recover the intrinsic and projected ring properties when the \hi\ disk is at intermediate inclination (generally in the range $40^\circ - 75^\circ$) and spatially resolved by $\sim8-10$ beams across the major axis (e.g.\ \citealt{bosma78}, \citealt{Kamphuis2015}).


More recent tilted-ring codes have generalized the approach for application directly to the 3D datacubes themselves (e.g.\ \textsc{TiRiFiC}, \citealt{Jorza2007}; \textsc{KinMS}, \citealt{Davis2013}; \textsc{3DBarolo}, \citealt{diTeodoro15}; \textsc{FAT}, \citealt{Kamphuis2015}; \textsc{GBKFit}, \citealt{Bekiaris2016}). 3D techniques have two main advantages relative to 2D ones: first, they allow for more complicated morphological and kinematic models to be applied to deep, high-resolution data \citep[e.g.][]{Jorza2009,Khoperskov2014,DiTeoDoro2021,Jorzsa2021}; and second, they can be robustly applied at lower spatial resolutions and across a wider range of disk geometries than in 2D \citep[e.g.][]{Kamphuis2015, diTeodoro15, Lewis2019, Jones2021}. Given the size distribution of sources implied by  Fig.~\ref{Fig:Size-SNPlot}, it is this latter property that makes 3D techniques most suitable for homogeneous modelling of PDR1 detections.

We work with \fat\ and  \barolo, two publicly-available codes designed to automatically apply 3D tilted-ring models to samples of \hi\ datacubes. Below, we describe the salient properties of both algorithms in the context of the PDR1 kinematic analysis.

\subsubsection{FAT}
\fat\footnote{\href{https://github.com/PeterKamphuis/FAT}{https://github.com/PeterKamphuis/FAT}} \citep{Kamphuis2015} automates the application of \tirific\footnote{\href{https://gigjozsa.github.io/tirific/}{https://gigjozsa.github.io/tirific/}} \citep{Jorza2007}, one of the first and most well-developed 3D tilted-ring codes.  \tirific\ constructs models by populating rings with tracer particles, projecting them into a 3D datacube, and convolving the result with a 3D kernel to match the spatial and spectral resolution of the data to which the model is compared. The model is then optimized by computing the channel-by-channel goodness of fit using an implementation of the `golden section' search algorithm \citep{Press1992}. 


The basic approach implemented in \fat\ is to automatically initialize \tirific\ using parameters determined from applying the \sofia\ source finder to the input datacube, and then to iteratively apply \tirific, usually with increasing complexity, until a satisfactory fit is achieved.  \fat\ begins with a flat-disk model in which the ring geometries are independent of galactocentric radius $R$, and has the functionality to explore radial variations in subsequent iterations or to fit flat-disk models. By design, \fat\ estimates an axisymmetric rotation curve but computes the surface brightness profile on the approaching and receding sides of the disk separately.  Once a satisfactory fit of the parameters is found, radial variations are smoothed by a polynomial fit to avoid artificial fluctuations from the \tirific\ fitting algorithm, with differences between smoothed and unsmoothed curves returned as uncertainties for some parameters.  

In a series of validation tests on real and simulated data, \citet{Kamphuis2015} show that \fat\ can reliably recover both the geometries and kinematics of \hi\ disks with inclinations ranging from $20^\circ - 90^\circ$ that are spatially resolved by at least 8 beams across their major axes, while extensive tests by \citet{Lewis2019} imply that \fat\ can recover inclinations and rotation curves for flat, axisymmetric disks resolved by as few as 3.5 beams across their \textcolor{black}{major axis diameters, $D_{HI}$,} in the inclination range $35^\circ - 80^\circ$. \textcolor{black}{We note that $D_{HI}$ differs from the SoFiA-returned \ellmaj\ shown in Fig.~\ref{Fig:Size-SNPlot} (see Sec.~\ref{Sec:CatalogueAndProducts})}.  We work with \fat\ version 2.01.

\subsubsection{3DBarolo}

\barolo\footnote{\href{https://editeodoro.github.io/Bbarolo/}{https://editeodoro.github.io/Bbarolo/}}  \citep{diTeodoro15} is a tilted-ring code that has been extensively used to apply 3D models to \hi\ datasets in different resolution and $S/N$ regimes. Many elements of the \barolo\ implementation are similar to those described for \fat\ above; below, we highlight differences that are relevant to the PDR1 kinematic analysis.

Key \barolo\ features that differ from \fat\ are parameter initialization, model optimization and flux normalization.  \barolo\ can use a built-in source finder based on \textsc{DUCHAMP} \citep{Whiting2012} to initialize the models, or the user can specify initial parameter estimates directly.  Once the source(s) are found, the model is optimized on a ring-by-ring basis using the Nelder-Mead algorithm \citep{Nelder65}, where beam effects are mimicked by convolving each velocity channel with a 2D kernel. \barolo\ can compute radial variations of the geometric parameters using a number of different strategies such as polynomial fits or Bezier interpolation, or can return median values if a flat-disk model is specified. The model cube flux is normalized in 2D using the observed moment 0 map, either on a pixel-by-pixel basis or on an azimuthal ring basis. 
This approach increases the efficiency of the \barolo\ optimization relative to the channel-by-channel method adopted in \tirific, but limits the range of disk inclinations and surface density distributions that can be robustly recovered.   \barolo\ implements a Monte Carlo approach to estimate uncertainties for some parameters, where models are varied around the best fit until the residuals increase by some factor (typically 5\%). 

In a series of validation tests on real data, \citet{diTeodoro15} show that \barolo\ can efficiently recover the geometries and kinematics of well-resolved and moderately-resolved \hi\ disks at intermediate inclinations from the THINGS \citep{Walter2008} and WHISP \citep{Swaters2002} surveys, respectively, while tests on both real data and galaxy mocks imply that \barolo\ can recover rotation curves and velocity dispersion profiles in systems resolved by as few as 2 beams along the major axis when the inclination is fixed and in the range $45^\circ - 75^\circ$. We work with \barolo\ version 1.6.

\subsection{Application to PDR1 Detections}
\label{Sec:ModelCodeComp}

The key differences between \fat\ and \barolo\ described above imply that the same fitting options applied to the same dataset by each code may yield different optimizations. These differences are typically small for spatially well-resolved, high S/N detections, but may be significant in the low-resolution, low-S/N regime in which most PDR1 sources lie (see Fig.~\ref{Fig:Size-SNPlot}; \citealt{Kamphuis2015}; \citealt{diTeodoro15}). Early in the pilot survey phase, we therefore explored a suite of different \fat\ and \barolo\ model applications to over a dozen Hydra TR1 detections in order to develop the technique we ultimately adopted. Because the sizes and S/N of most PDR1 sources pose challenges to tilted-ring modelling even with 3D applications, we restricted the analysis to simple, flat-disk models where the disk geometry does not vary with $R$. 

\begin{figure*}
\centering
    \includegraphics[width=0.7\textwidth]{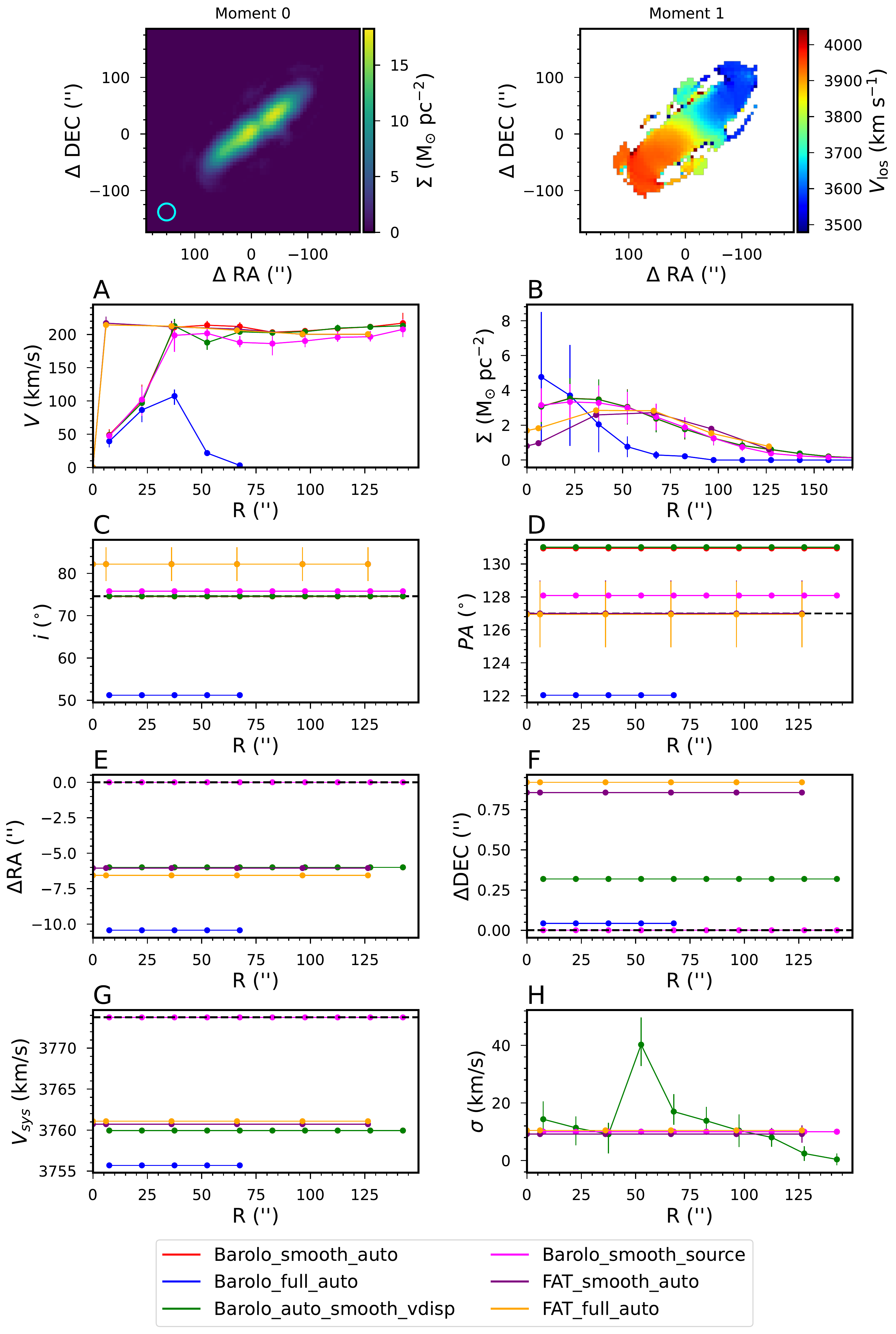} \caption{Comparison between different flat-disk model outputs applied with \fat\ and \barolo\ to WALLABY J103915-301757 (\ellmaj$=3.9~\textrm{beams}$, $\log(S/N_{obs})=1.5$). The top row shows the moment 0 and moment 1 maps of this source, with the cyan circle indicating the size of the beam.  The panels below show plots of the rotation curve (A), surface density profile (B), inclination (C), position angle  (D), kinematic centre \textcolor{black}{relative to the PDR1 source centroid} (E and F), systemic velocity (G) and velocity dispersion profile (H) as a function of galactocentric radius $R$ for the flat-disk models given in the legend (see text for details), evaluated at the locations given by the points.  The dashed black lines in panels C, D, E, F and G indicate the PDR1 source parameters for those quantities from \citetalias{SoFiARelease}. The error bars on some profiles in some panels are the final uncertainties returned by either \fat\ or \barolo\ for that model application. }
  \label{Fig:ModelCompPlot}
\end{figure*}



  
  This experimentation revealed that among possible flat-disk modelling choices in \fat\ and \barolo, a) the $S/N$ of the detected emission in each channel and b) the model parameter initialization can both strongly influence the optimizations returned in the PDR1 regime, with variations in other algorithm switches having comparatively  minor effects. The $S/N$ of the emission per channel can impact the reliability of the built-in source finder which initializes parameters, hence the importance of that modelling choice. The parameter initialization, in turn, can impact the model outputs because optimization schemes such as Golden Section (as in \fat) and Nelder-Mead (as in \barolo) require robust initial guesses to converge in the complex, multi-dimensional parameter spaces characteristic of 3D tilted-ring models (e.g.~\citealt{Bekiaris2016}).
  
 We illustrate these trends in the PDR1 regime in Fig.~\ref{Fig:ModelCompPlot}, which shows the output parameters for several flat-disk \textcolor{black}{models} applied to WALLABY J103915-301757 (\ellmaj $=3.9~\textrm{beams}$, $\log(S/N_{obs})=1.5$) with \fat\ and \barolo.   This example is just one of the dozen galaxies tested with different modelling options and illustrates well how different choices affect the resulting models.  The main differences between the different fitting attempts are the spectral resolution of the cubelet to which the model is applied (either the full-resolution cubelet, or a 3-channel Hanning-smoothed cubelet), and the choice of geometric parameter initialization (either initialized automatically by the code or initialized by the user to the PDR1 source values from \citetalias{SoFiARelease}):
 \begin{itemize}
    \item \textbf{Barolo\_full\_auto}: \barolo\ applied to the full-resolution cubelet, with automated parameter initialization;
    \item \textbf{Barolo\_smooth\_auto}: \barolo\ applied to the spectrally-smoothed cubelet, with automated parameter initialization; 
    \item \textbf{Barolo\_smooth\_source}: \barolo\ applied to the spectrally-smoothed cubelet, with geometric parameters initialized to the PDR1 source values; 
    \item \textbf{FAT\_full\_auto}: \fat\ applied to the full-resolution cubelet, with automated parameter initialization;
    \item \textbf{FAT\_smooth\_auto}: \fat\ applied to the spectrally-smoothed cubelet, with automated parameter initialization; 
    \item \textbf{Barolo\_auto\_smooth\_vdisp}: \barolo\ applied to the spectrally-smoothed cubelet, allowing the velocity dispersion to vary.
\end{itemize}
We note that since \fat\ does not allow the user to initialize parameters, there no such model with this option in Fig.~\ref{Fig:ModelCompPlot}. We note also that since many PDR1 detections have no optical counterparts (particularly in Norma TR1, which is close to the Galactic Plane), we do not attempt to initialize geometric parameters with photometric values. We note that the Barolo\_auto\_smooth\_vdisp shown in Fig. \ref{Fig:ModelCompPlot} involves a different fitting mode than the other models, and is discussed further below.


 Comparing the optimized rotation curves (Fig.~\ref{Fig:ModelCompPlot}A), surface density profiles (Fig.~\ref{Fig:ModelCompPlot}B) and disk geometries (Fig.~\ref{Fig:ModelCompPlot}C--G) across models for WALLABY J103915-301757, the \barolo\ application to the full-resolution cubelet (Barolo\_full\_auto) differs markedly from the other outputs: its radial extent is much smaller than that of the source plotted in the top row, and the disk geometry (most notably the inclination) is discrepant with the source morphology. This model failure stems from an incorrect source identification and parameter initialization by the \barolo\ source finder due to the low S/N of the emission in each channel of the full-resolution cubelet. 

\textcolor{black}{Regardless of the model, the position angle and geometric center (Fig.~\ref{Fig:ModelCompPlot}D--F) are recovered well.  Given the pixel size is $\sim 6$\arcsec, the kinematic center is recovered within less than 2 pixels.  The successful measurement of these three geometric parameters is typical for kinematic modelling as they tend to have fewer degeneracies with other parameters than the inclination or systemic velocity. If there are large differences between the kinematic center and position angle for various fits, it indicates a failure in one or more of those fits.  The systemic velocity itself (Fig.~\ref{Fig:ModelCompPlot}G) shows a larger variation, but for the smoothed cubes, the differences are only of the order of 1-2 channels. The greatest outlier is the Barolo\_smooth\_source model, which also is an outlier in terms of the spatial center.}

 We find that in general, models applied smoothed to cubelets are more stable and converge faster than those applied to the full-resolution cubelets, with little difference between the optimized values when both models succeed (e.g. FAT\_full\_auto and FAT\_smooth\_auto in Fig.~\ref{Fig:ModelCompPlot}). This is not unexpected since the modelled PDR1 detections are spectrally well-resolved in both the full-resolution and smoothed cubelets, while the per-channel S/N of the emission is $\sim$50\% higher in the latter. We therefore elect to kinematically model PDR1 cubelets that have been spectrally smoothed by a 3-channel Hanning window. 
 
 Another trend that emerges among successful models in Fig.~\ref{Fig:ModelCompPlot} is that the outputs from different models applied using the same code (e.g.\ FAT\_full\_auto vs.\ FAT\_smooth\_auto) are typically more similar than those from the same model applied by different codes (e.g.\ Barolo\_full\_smooth vs.\ FAT\_full\_smooth), with differences that can well exceed the returned uncertainties. We find this to be generally the case for the rotation curve and surface brightness profiles at radii within $\sim$1 beam of the kinematic centre (Fig.~\ref{Fig:ModelCompPlot}A and B): the \fat-returned rotation curves tend to rise more steeply and surface brightness distributions tend to exhibit greater central depressions than the \barolo\ counterparts, particularly for relatively high inclination and/or poorly spatially resolved sources.  These discrepancies may stem in part from the different radial grid definitions adopted by the codes (\barolo\ returns model values at the ring mid-points, whereas \fat\ uses the ring edges), but differences in optimization methodology (see Sec.~\ref{sec:TRModelling}) likely play a stronger role.  Regardless of the cause, the key point is that the differences between successful \fat\ and \barolo\ fits are typically larger than the reported uncertainties. Our PDR1 modelling approach therefore adopts an average of the models returned by each code as the optimal model, and differences between them as a measure of uncertainty.

 
 \begin{figure}[t]
\centering
    \includegraphics[width=\textwidth]{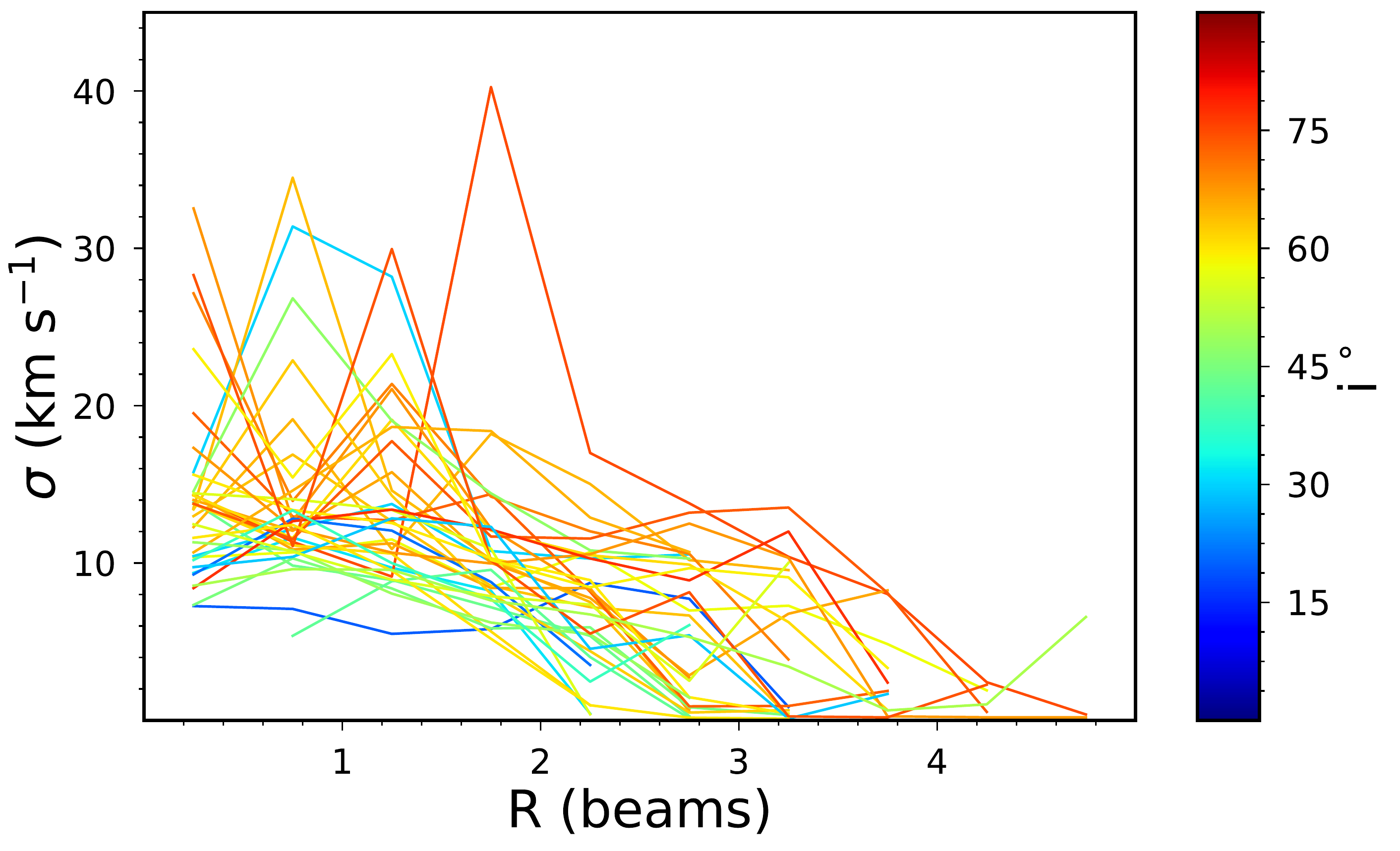} \caption{Velocity dispersion profiles from models identical to Barolo\_auto\_smooth\_vdisp in Fig.~\ref{Fig:ModelCompPlot}, applied to all 36 PDR1 sources with $2\le \ellmaj \le 4 $ and $1.25\le \log(S/N_{obs})\le 1.5$. The profiles are coloured according to the model disk inclination.}
  \label{Fig:VelDispPlot}
\end{figure}
 
 The dashed horizontal lines in  Fig.~\ref{Fig:ModelCompPlot} plot the PDR1 source parameters that best approximate the geometric parameters returned by the kinematic modelling: $\cos^{-1}(\code{ell\_min} / \code{ell\_maj})$,  \code{kin\_pa}, \code{ra}, \code{dec}, and \code{freq} (see table~3 of \citetalias{SoFiARelease}) converted to the appropriate units are shown in Fig.~\ref{Fig:ModelCompPlot}C-G respectively. Model Barolo\_smooth\_source initializes the model geometric parameters to these values; for most successful models the output parameters are nearly identical to the inputs, and different from the outputs from runs in which the geometric parameters are initialized automatically in either the Barolo\_smooth\_auto model or in the FAT\_smooth\_auto model. We find this to be generally true for the PDR1 models, and speculate that the tilted-ring parameter space is sufficiently complex that the \barolo\ optimizer is unlikely to find a step that improves the goodness of fit during the runs as configured. Since the PDR1 parameters only approximate the kinematic model parameters in the first place, we elect to use automatic source initialization in the kinematic analysis.

We now discuss model Barolo\_auto\_smooth\_vdisp in Fig.~\ref{Fig:ModelCompPlot}, which is identical to Barolo\_auto\_smooth except that the disk velocity dispersion is allowed to vary with $R$. Save for small differences between the rotation curves at $R \sim 50$\arcsec\ and the very different velocity dispersion profiles, the returned parameters are almost identical between the two models, with corresponding lines overlapping completely in Fig.~\ref{Fig:ModelCompPlot}B--G. 

 We find the independence of the model outputs on the velocity dispersion switch, as well as the large radial variations in velocity dispersion when it is allowed to vary, to be general trends in our PDR1 models. The velocity dispersion variations are further illustrated in Fig.~\ref{Fig:VelDispPlot}, where models identical to Barolo\_auto\_smooth\_vdisp were applied to the 36 PDR1 sources \textcolor{black}{in Hydra TR2, Norma TR1, and NGC 4636 TR1} with $2 \le \code{ell\_maj} \le 4 $ and $1.25\le \log(S/N_{obs})\le 1.5$. This figure shows that, independent of disk inclination, there is a general trend of decreasing velocity dispersion with increasing $R$, but also variations across profiles and between them that well exceed the relatively tight range $8\,\kms \le \sigma \le 12 \,\kms$ measured from high-resolution \hi\ maps \citep{Tamburro2009}. This is perhaps not surprising given the $\sim\,12 \kms$ resolution of the Hanning-smoothed cubelets that we model. We therefore keep the velocity dispersion fixed to $\sigma = 10\,\kms$ (intermediate to the range of values typically measured) in the PDR1 kinematic models.

\section{The WALLABY Kinematic Analysis Proto-Pipeline}
\label{Sec:ModelProcedure}

\begin{figure*}
\centering
    \includegraphics[width=170mm]{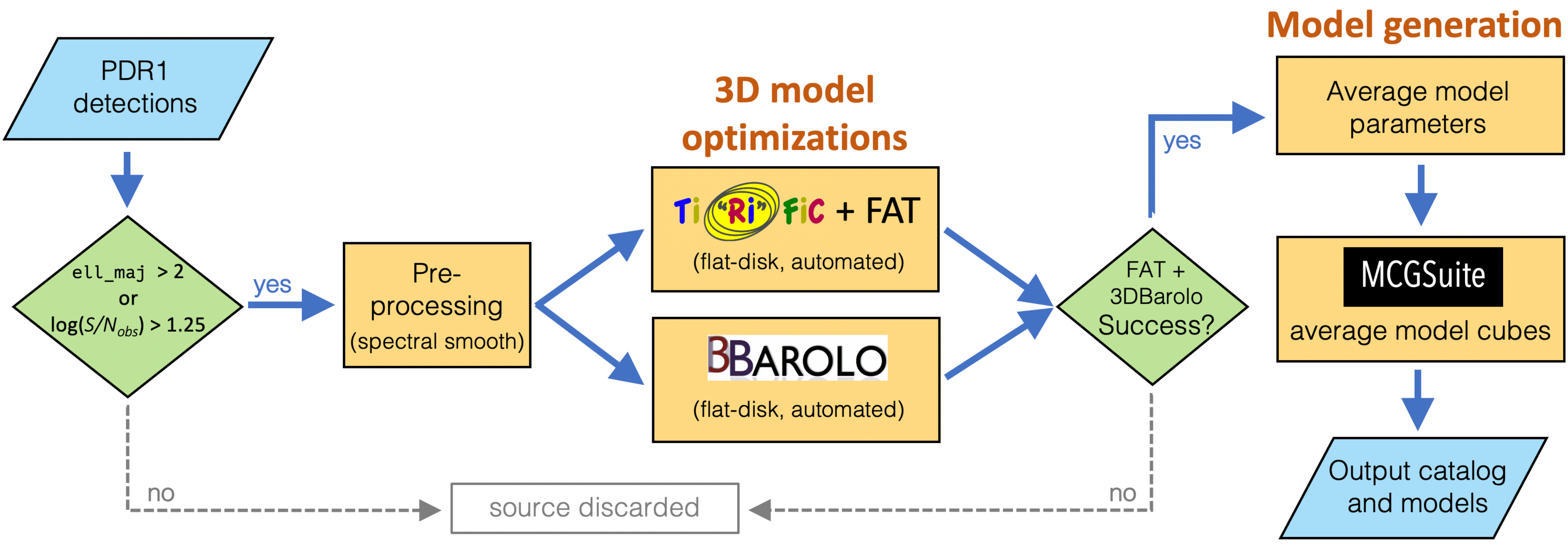} \caption{A schematic of the \textsc{WKAPP} modelling process.  The blue parallelograms indicate data products, the green diamonds indicate decision points, and yellow boxes indicate automated code.}
  \label{Fig:FlowDiagram}
\end{figure*}

Having explored some key considerations for kinematic modelling of PDR1 detections in Sec.~\ref{sec:TRModelling}, we now describe the approach adopted to derive flat-disk tilted-ring models by applying \fat\ and \barolo\ to pre-processed PDR1 cubelets and averaging successful fits. We call the procedure the WALLABY Kinematic Analysis Proto-Pipeline (\textsc{WKAPP}), and the full set of driving scripts is available from its distribution page\footnote{\href{https://github.com/CIRADA-Tools/WKAPP}{https://github.com/CIRADA-Tools/WKAPP}}. Fig.~\ref{Fig:FlowDiagram} summarizes the modelling steps:
 \begin{enumerate}
    \item Select detections on which kinematic modelling is attempted and pre-process their PDR1 cubelets;
    \item Apply \fat\ and \barolo\ models to pre-processed cubelets;
    \item Generate optimized models and uncertainties by averaging successful model fits;
    \item Compute surface density profiles from PDR1 source Moment~0 maps using optimized model geometries.
\end{enumerate}
 
We describe each of these steps in Sec.~\ref{subsec:gal_selection}--\ref{subsec:mod_surfdens}, the model outputs in Sec.~\ref{subsec:mod_outputs}, and some limitations of the current approach in Sec.~\ref{subsec:mod_lim}.

\subsection{Detection selection and cubelet pre-processing}
\label{subsec:gal_selection}

The first step of \textsc{WKAPP} is to select a set of PDR1 detections on which kinematic modelling is attempted. Validation tests on \fat\ and \barolo\ suggest that the algorithms can be successfully applied to \hi\ disks with diameters $D_{HI}$ that are resolved by as few as 2-3 beams depending in the $S/N$ \citep{diTeodoro15, Lewis2019}.  We use the PDR1 size measure \ellmaj\ in our selection, which is typically a factor of two smaller than $D_{\hi}$ in our successful models (see Sec. \ref{Sec:Populations} for a comparison of $D_{\hi}$ to \ellmaj).  Therefore we attempt to model all detections with $\ellmaj \ge 2$ beams.  Because \ellmaj\ is not a direct measure of disk size, we also attempt to model all detections with $\log(S/N_{obs})\ge 1.25$, even if they are below the size threshold. These selection cuts result in 209 unique PDR1 detections that we attempt to model, shown by the red and blue points in Fig.~\ref{Fig:Size-SNPlot}. 


Next, the PDR1 cubelets selected for modelling are pre-processed in two steps.  First, the spectral axis of the cubelets is converted to velocity units from the  frequency units provided in the PDR1 data release.
Second, the cubelets are Hanning-smoothed by three spectral channels (to a resolution of $11.7\,\kms$ at $z=0$) using \barolo. As discussed in Sec.~\ref{Sec:ModelCodeComp}, the main driver of this choice is an increase in model stability for essentially the same model fit quality.  It also decreases the \fat\ and \barolo\ run time since there are fewer spectral channels.

\subsection{Application of \fat\ and \barolo\ models}
\label{subsec:mod_application}

For each of the PDR1 detections selected as described above,  we automatically apply flat-disk tilted-ring models to the pre-processed cubelets using \barolo\ and \fat. As discussed in Sec. \ref{Sec:ModelCodeComp}, we allow each code to automatically initialize all parameters, and we fix the velocity dispersion to $10\,\kms$ in the models.  

For \barolo, the ring widths are set to 2 rings/beam and we use the \code{SEARCH} source-finding method and the azimuthal normalization method (in order to be as similar to \fat\ as possible).  \fat\ does not allow the ring size to be specified, but it generally fits objects with 2 rings/beam as well.  For completeness, both the input and results files from the \barolo\ and \fat\ applications to each successfully-modelled detection are distributed with the data release (see Sec.~\ref{Sec:CatalogueAndProducts}). 



\subsection{Fit Inspection and Optimal Model Geometry and Rotation Curve}
\label{subsec:mod_optimal}

Only a subset of selected sources are successfully modelled using either \fat\ or \barolo: some show complicated structures that are not well-described by flat disks, some are actually pairs of galaxies, and many have too low of a resolution or $S/N$ to be successfully modelled (see Sec.~\ref{Sec:CatalogueAndProducts}).  The results of the \barolo\ and \fat\ fits for each source are therefore visually examined to determine their success.  If either code fails to produce a model, if the final model for either code is non-physical (for example, Barolo\_full\_auto in Fig.~\ref{Fig:ModelCompPlot}; see Sec.~\ref{Sec:ModelCodeComp}), or if the models returned differ strongly (for example, $\delta \sin(i) > 0.2$ between the \fat\ and \barolo\ results), then the source is discarded from the kinematic modelling sample (see Fig.~\ref{Fig:FlowDiagram}). 


If both the \barolo\ and \fat\ fits are successful, then the two fits are averaged together in three distinct steps to generate an optimal kinematic model. The first is to directly average the geometric parameters (center, $V_{sys}$, inclination, and position angle), with the uncertainty set to half the difference between them.    Table \ref{Tab:GeoAvg} shows an example for WALLABY J163924-565221 (\code{ell\_maj}$=4.5$ beams and $\log(S/N_{\mathrm{obs}})=1.53$), a relatively large and high $S/N$ PDR1 detection (see Fig.~\ref{Fig:Size-SNPlot}).



The averaged model geometry is then used to calculate the optimal rotation curve from the outputs of the \fat\ and \barolo\ models. Since these models typically have different radial extents and are evaluated at different values of $R$, a final radial grid must be constructed.  The final grid has two points per beam; the innermost point is set to the larger of the two smallest model $R$, which also defines the grid values. To optimize the radial extent of the models, the outermost rotation curve point is the largest $R$ on the grid at which one model is interpolated and the other is extrapolated by no more than half a beam. Figure~\ref{Fig:AveragingProcess} shows an example of the radial grid definition for the rotation curve of WALLABY J163924-565221.



With the final geometric parameters calculated and the radial grid set, the \barolo\ and \fat\ rotation curves are adjusted to the final inclination and interpolated onto the final grid using a spline fit.  As with the geometric parameters, the uncertainty on each rotation curve point is set to half the difference between the two interpolated curves at that $R$. We also propagate the effect of the inclination uncertainty to the rotation curve, providing a separate value for this source of error.  It is recommended that these two uncertainties be added in quadrature when \textcolor{black}{working} with the model rotation curves.  An example of the optimal rotation curve calculation is given in Fig. \ref{Fig:AveragingProcess} for WALLABY J163924-565221. 

\textcolor{black}{We note that the \fat\ and \barolo\ model rotation curves are generated with a degree of internal consistency.  But it is not guaranteed that our optimized rotation curves will have similar levels of self-consistency as they are generated by averaging the interpolatated, inclination corrected \fat\ and \barolo\ outputs.  However, the visual inspection of the different fits as well as a final examination of the optimized model help to avoid any inconsistencies, and in practice the best fitting disk centres and position angles are typically very similar (see Sec.~\ref{Sec:ModelCodeComp}). We therefore judge that the successful models have rotation curves that are internally consistent with the rest of the model parameters.}



\subsection{Surface Density Profile Computation}
\label{subsec:mod_surfdens}

In the final WKAPP step, the surface density profile is calculated from ellipse fits to the PDR1 detection Moment 0 map and the average geometry. In other words, the surface density profile is derived separately from the \fat\ and \barolo\ estimates of this parameter, but using the same disk geometry as in the optimized model. This approach is similar to the \barolo\ procedure for calculating surface densities but differs strongly from  the \fat\ approach (see Sec.~\ref{subsec:TRmodcodes}), where they are constrained directly from the cube; since the \fat\ approach has not been vetted in the resolution and $S/N$ regime of the PDR1 detections, we use ellipse fits for this first public data release with the goal of using cube fits in future ones (see Sec.~\ref{subsec:mod_lim}).

The optimized surface density profile is computed using the same radial grid values as the rotation curve, but with the extent determined by the PDR1 mask width along the kinematic major axis of the Moment 0 map. In practice, this implies that the surface density profile of a given model typically extends to larger $R$ than its rotation curve; this choice implies that the majority of the surface density profiles extend to the characteristic density $\Sigma = 1\,\mathrm{M_\odot \, pc^{-2}}$ at which disk radii are typically defined \citep[e.g.][]{Wang16}, although this requires extrapolating the disk geometry beyond the region used in the model fits. We adopt the standard error on the mean as the uncertainty in the measured profiles, that is the standard deviation of the pixels in each ring divided by the square root of the number of beams in that ring.

In addition to providing the surface density profile directly measured from the ellipse fits, we also provide a version to which a standard $\cos(i)$ correction has been applied to deproject the profiles to a face-on view. We caution that this correction can strongly under-estimate the inner surface densities of marginally-resolved HI disks, as is the case for many PDR1 detections (see Fig.~\ref{Fig:Size-SNPlot}). In addition, we do not attempt to correct the outer surface density profiles for beam smearing effects. We discuss both effects in Sec.~\ref{subsec:mod_lim}, and recommend that their impact be considered when using the corrected surface density profiles.

\subsection{Model Outputs}
\label{subsec:mod_outputs}


\begin{table}[t]
    \centering
    \begin{tabular}{|c|c|c|c|c|c|}
        \hline 
         Parameter & Units & \fat\  & \barolo\  & Model & Uncertainty\\
         \hline 
         $X$ & px & 40.5 & 39.5 & 40.0 & 0.5\\ 
         $Y$ & px & 33.5 & 33.8 & 33.7 & 0.1\\ 
         $V_{sys}$ & $\kms$ & 1466.9 & 1465.7 & 1466.3 & 0.6\\ 
         Inc & deg & 49.4 & 37.0 & 43.2 & 6.2\\ 
         PA & deg & 249.9 & 245.6 & 247.8 & 2.2\\ 
         \hline 
    \end{tabular}
    \caption{Geometric parameter averaging for the WKAPP model to WALLABY J163924-565221 (\code{ell\_maj}$=4.5$ beams and $\log(S/N_{\mathrm{obs}})=1.53$).  The \fat\ and \barolo\ columns show the results from the fits to the galaxy using the respective codes. the Model and Uncertainty columns show the average geometric parameters and rounded uncertainties adopted.}
    \label{Tab:GeoAvg}
\end{table}

Every PDR1 source that is successfully modelled by \textsc{WKAPP} is characterized by a set of model parameters as listed in Table \ref{Tab:KinematicParameters}.  The geometric parameters and associated uncertainties are single values, while the rotation curve and surface density profiles and their associated uncertainties are arrays. 

We note that among the geometric parameters provided for each model are PAs in pixel coordinates (\code{PA\_model}) and in global equatorial coordinates (\code{PA\_model\_g}). For most PDR1 detections, there is a small but non-zero rotational offset between those two coordinate systems that is defined by the PDR1 cubelet header. This results in a small but systematic difference between \code{PA\_model} and \code{PA\_model\_g} (typically less than 2 degrees).

\begin{figure}[t]
\centering
    \includegraphics[width=\textwidth]{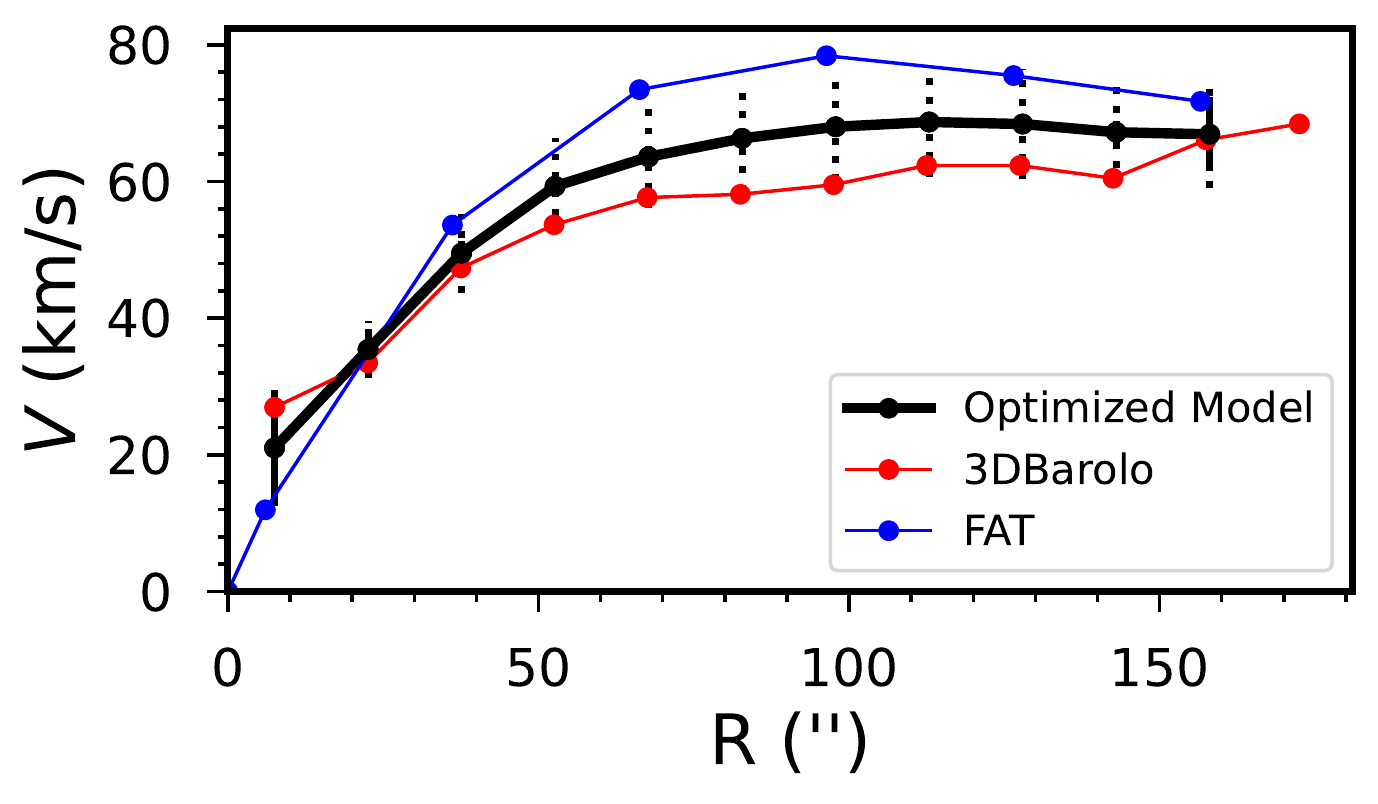} \caption{Example showing how the \barolo\ and \fat\ rotation curve fits are combined into a single average model for WALLABY J163924-565221, where the geometric parameters are given in Table~\ref{Tab:GeoAvg}.  The black line shows the optimal rotation curve, while the red and blue lines show the outputs from the automated \barolo\ and \fat\ models fits respectively.  The solid error bars show the uncertainty from averaging the interpolated inclination-adjusted rotation curves, while the dashed error bars show the uncertainty in the inclination propagated into the rotation curve. In this example, the latter uncertainty is much larger than the former for most points.}
  \label{Fig:AveragingProcess}
\end{figure}

As described in Sec.~\ref{subsec:mod_optimal}, we provide estimates of uncertainty from two different sources for each rotation curve: the first (\code{e\_Vrot\_model}) arises from the \fat\ and \barolo\ averaging process, and the second (\code{e\_Vrot\_model\_inc}) is the contribution to the uncertainty on the rotation curve obtained by propagating the uncertainty on the inclination (\code{e\_Inc\_model}). Fig.~\ref{Fig:AveragingProcess} provides an example of these two sources of uncertainty for WALLABY J163924-565221. We recommend adding these sources in quadrature when \textcolor{black}{using} the rotation curve. 

We also provide estimates of the uncertainty on the surface density profile from two sources. The first (\code{e\_SD\_model}) is the standard error of the pixels in each ring, which we recommend be adopted as the uncertainty in the projected surface density profile (\code{SD\_model}). We also provide an estimate of the statistical uncertainty for the profile deprojection (\code{e\_SD\_FO\_model\_inc}) obtained by propagating the uncertainty on the inclination (\code{e\_Inc\_model}). We recommend adding these sources in quadrature when the deprojected surface density profile (\code{SD\_FO\_model}) is used for scientific analysis, but caution that for many PDR1 sources systematic errors in the standard $\cos(i)$ correction dominate. We discuss this further in Sec.~\ref{subsec:mod_lim} below.   


We also provide a quality flag for each model:
\begin{itemize}
    \item \code{QFlag\_model} = 0: No obvious issues.
    \item \code{QFlag\_model} = 1: $\code{Inc\_model} \leq 20^{\circ}$ or $\code{Inc\_model} \geq 80^{\circ}$.
    \item \code{QFlag\_model} = 2: $\code{e\_Vsys\_model} \geq 15\,\kms$. 
    \item \code{QFlag\_model} = 3: Both conditions 1 and 2 are met. 
\end{itemize}

\begin{table*}
    \centering
    \begin{tabular}{|p{0.13\textwidth}|>{\centering}p{0.1\textwidth}|p{0.07\textwidth}|>{\centering\arraybackslash}p{0.5\textwidth}|}
         \hline
         Name & Type & Unit & Description \\
         \hline 
          \code{X\_model} & double & px & x-coordinate of the kinematic center$^{\dagger}$\\ 
          \code{e\_X\_model} & double & px & Uncertainty in \code{X\_model}$^{\dagger}$\\
          \code{Y\_model} & double & px & y-coordinate of the kinematic center.$^{\dagger}$\\ 
          \code{e\_Y\_model} & double & px & Uncertainty in \code{Y\_model}$^{\dagger}$\\
          \code{RA\_model} & double & deg & Right ascension (J2000) of the kinematic center \\
          \code{e\_RA\_model} & double & deg & Uncertainty in \code{RA\_model}$^{\dagger}$\\
          \code{DEC\_model} & double & deg & Declination (J2000) of the kinematic center \\
          \code{e\_DEC\_model} & double & deg & Uncertainty in \code{DEC\_model}$^{\dagger}$\\
          \code{Vsys\_model} & double & $\kms$ & Heliocentric systemic velocity\\ 
          \code{e\_Vsys\_model} & double & $\kms$ & Uncertainty in \code{Vsys\_model}\\
          \code{Inc\_model} & double & deg &  Inclination\\ 
          \code{e\_Inc\_model} & double & deg & Uncertainty in \code{Inc\_model}\\
          \code{PA\_model} & double & deg & Position angle in pixel coordinates (counterclockwise from x=0)$^{\dagger}$\\ 
          \code{e\_PA\_model} & double & deg & Uncertainty in \code{PA\_model}$^{\dagger}$\\
          \code{PA\_model\_g} & double & deg & Position angle in equatorial coordinates (East of North)\\ 
          \code{e\_PA\_model\_g} & double & deg & Uncertainty in \code{PA\_model\_g}\\
          \code{Rad} & double array array & arcsec & Radial grid for \code{Vrot\_model}\\
          \code{Vrot\_model} & double array & $\kms$ & Rotation curve \\
          \code{e\_Vrot\_model} & double array & $\kms$ & Uncertainty in \code{Vrot\_model} from the averaging process\\
          \code{e\_Vrot\_model\_inc} & double array & $\kms$ & Uncertainty in \code{Vrot\_model} due to \code{e\_Inc\_model}\\
          \code{Rad\_SD} & double array & arcsec & Radial grid for \code{SD\_model} and \code{SD\_FO\_model}\\
          \code{SD\_model} & double array & M$_{\odot}$ pc$^{-2}$& Projected surface density profile\\
          \code{e\_SD\_model} & double array & M$_{\odot}$ pc$^{-2}$&  Uncertainty in \code{SD\_model}\\
          \code{SD\_FO\_model} & double array & M$_{\odot}$ pc$^{-2} $& Deprojected surface density profile using a $\cos(\code{Inc\_model})$ correction\\
          \code{e\_SD\_FO\_model\_inc} & double array & M$_{\odot}$ pc$^{-2}$ & The uncertainty in \code{SD\_FO\_model} due to \code{e\_Inc\_model}\\
          \code{QFlag\_model} & integer & & Kinematic model quality flag\\
          \hline 
          \multicolumn{4}{l}{$^{\dagger}\,$In pixel coodinates relative to the preprocessed cubelet\textcolor{black}{, which starts from the point (1,1)}.}
    \end{tabular}
    \caption{WKAPP model parameters.}
    \label{Tab:KinematicParameters}
\end{table*}

 We flag models with inclinations below $20^\circ$ and above $80^\circ$ (\code{QFlag\_model} = 1) because, although we judge these fits to be successful, these inclinations lie in the range where neither \fat\ nor \barolo\ have been vetted \citep{diTeodoro15, Kamphuis2015, Lewis2019}. We similarly have judged fits with $\code{e\_Vsys\_model} \geq 15\,\kms$ (\code{QFlag\_model} = 2) to be successful, but they are strong outliers in the distribution of this value (see Sec.~\ref{Sec:CatalogueAndProducts}) which may indicate a subtle failure that is not obvious through visual inspection. 
$\sim12\% $ (16/124) of all modelled sources have been flagged: \code{QFlag\_model} = 2 and \code{QFlag\_model} = 3 have each been assigned once, with the remaining 14 sources having been assigned  \code{QFlag\_model} = 1. 


\begin{table*}
    \centering
    \begin{tabular}{|p{0.2\textwidth}|>{\centering\arraybackslash}p{0.12\textwidth}|>{\centering\arraybackslash}p{0.42\textwidth}|}
         \hline
          File suffix & Type & Description \\
         \hline 
           \_AvgMod.txt & ascii file &    Model parameters\\
            \_DiagnosticPlot.png & PNG file &   Model summary plot  \\ 
           \_ProcData.fits & FITS cube &   Pre-processed cubelet \\ 
           \_ModCube.fits & FITS cube &  Model realization with pre-processed cubelet properties\\ 
          \_DiffCube.fits & FITS cube &  Data - model cube with pre-processed cubelet properties \\ 
            \_ModRotCurve.fits & FITS binary table &   Table containing the model rotation curve parameters  \\ 
           \_ModSurfDens.fits & FITS binary table &   Table containing the model surface density parameters  \\
            \_ModGeo.fits & FITS binary table &   Table containing the model geometry parameters  \\
            \_FullResProcData.fits & FITS cube &   Full spectral resolution cubelet with velocity units \\ 
            \_FullResModelCube.fits & FITS cube &  Model realization with full resolution cubelet properties\\
           \_FATInput.txt  & ascii file &   The input file of the \fat\ run\\ 
            \_FATMod.txt  & ascii file &   The results of the \fat\ run\\
           \_BaroloInput.txt & ascii file &   The input file of the \barolo\ run\\ 
           \_BaroloMod.txt  & ascii file &   The geometry and rotation curve results of the \barolo\ run\\
           \_BaroloSurfDens.txt  & ascii file &   The surface density results of the \barolo\ run\\
          \hline 
    \end{tabular}
    \caption{WKAPP data products available for each successfullly modelled PDR1 source.}
    \label{Tab:DataProducts}
\end{table*}

In addition to the catalog of model parameters for all kinematically modelled PDR1 sources, \textsc{WKAPP} also produces a number of data products for each source. They are listed in Table \ref{Tab:DataProducts}. Several products serve to group model parameters for individual sources into distinct files for ease of access: files with suffix  \code{\_AvgMod.txt} contain all model parameters for the PDR1 source in the prefix, while those with suffixes \code{\_ModRotCurve.fits}, \code{\_ModSurfDens.fits}, and \code{\_ModGeo.fits} store the rotation curve, surface density profile, and geometric parameters as FITS binary tables respectively.  Additionally, text files with \code{FAT} or \code{Barolo} in the suffix provide the input and output files from the automated \fat\ and \barolo\ applications described in Sec.~\ref{subsec:mod_application}. 

Several data and model cubelets are also provided as data products. The model cubelets are realizations of the optimized models using the stand-alone tilted-ring model generator \mcgsuite\ code\footnote{\href{https://github.com/CIRADA-Tools/MCGSuite}{https://github.com/CIRADA-Tools/MCGSuite}} (\citealt{Lewis2019}, Spekkens et al.\ in prep).  The pre-processed PDR1 cubelets to which \fat\ and \barolo\ are applied (see Sec.~\ref{subsec:mod_application}) are in files with suffix \code{\_ProcData.fits}. Realizations of the optimized models in cubelets with the same properties as the pre-processed data cubelets as well as data -- model cubelets are in files with suffixes \code{\_ModeCube.fits} and \code{\_DiffCube.fits}, respectively. For completeness, we also provide PDR1 cubelets at full spectral resolution with the frequency axis in velocity units (suffix \code{\_FullResProcData.fits}), as well as model realizations with the properties of those cubelets (suffix \code{\_FullResModelCube.fits}). 

\begin{figure*}
\centering
    \includegraphics[width=0.8\textwidth]{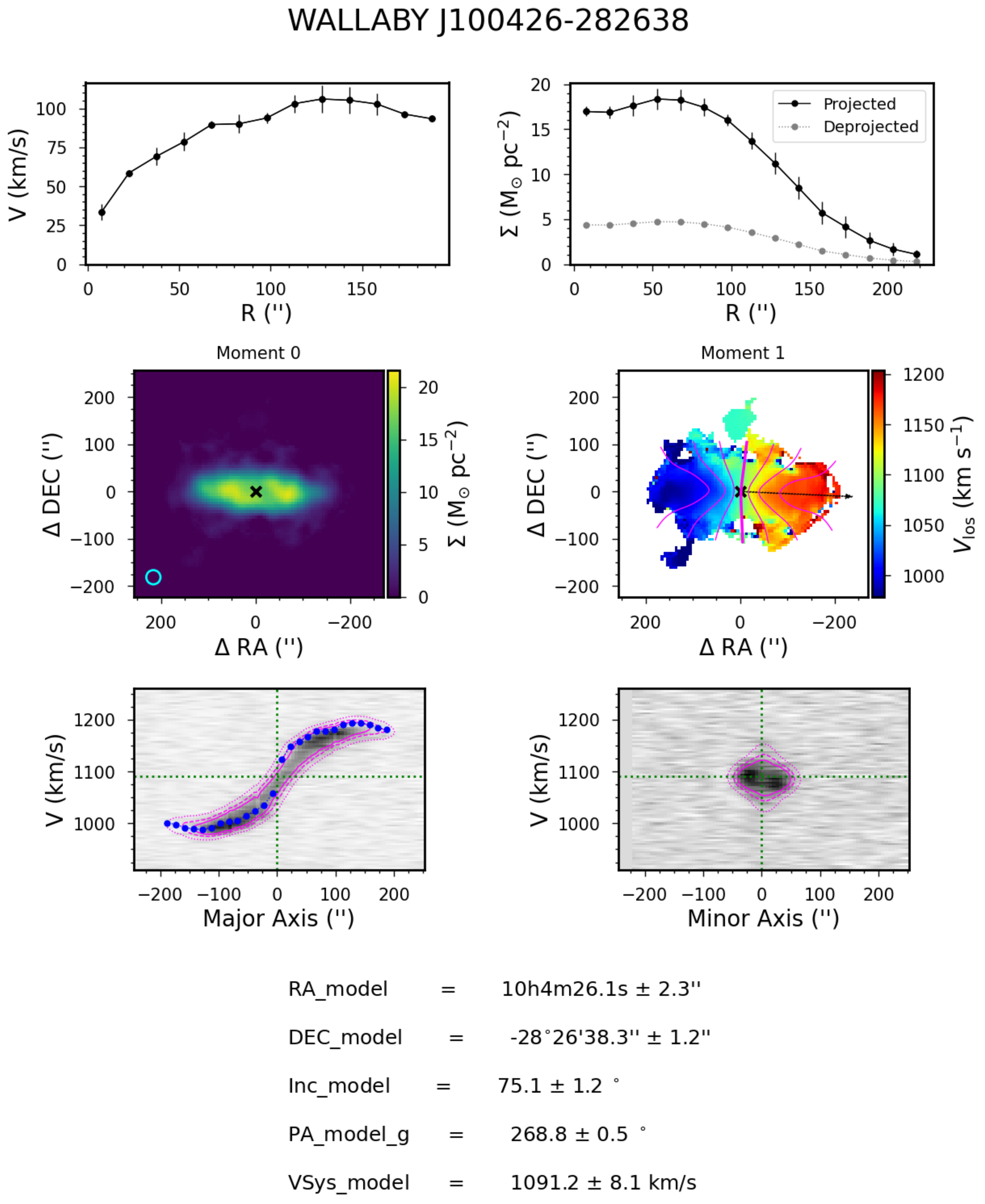} \caption{
    Sample \textsc{WKAPP} model summary plot for WALLABY J100426-282638 (\code{ell\_maj}$=5.0$ beams and $\log(S/N_{\mathrm{obs}})=1.83$).  Similar summary plots are included in the data release for each modelled PDR1 detection.  The upper left and right panels show the rotation curve and surface density profile of the optimized model.  The middle left panel shows the PDR1 Moment 0 map and the location of the model center marked with a black X.  The middle right panel shows the PDR1 Moment 1 map, along with the model velocity contours (constructed from the \mcgsuite\ cube realization), and the direction of the model position angle marked by a black arrow.  The bottom panels show the major and minor axis position-velocity (PV) diagrams (left and right panels respectively) along with the corresponding model PV diagrams (magenta lines).  The model contours are at 3 and 5$\sigma$ of the PV diagram noise.  The major axis PV diagram also shows the projected rotation profile ($\,=\, $\code{Vrot\_model}$\times\sin[$\code{Inc\_model}$]$).}
  \label{Fig:Summary}
\end{figure*}

Finally, a summary plot is provided for each modelled source as a PNG file with suffix \code{\_DiagnosticPlot.png}.  As an example, Figure \ref{Fig:Summary} shows the summary plot for WALLABY J100426-282638  (\code{ell\_maj}$=5.0$ beams and $\log(S/N_{\mathrm{obs}})=1.83$), one of the largest and highest $S/N$ PDR1 detections (see Fig.~\ref{Fig:Size-SNPlot}).  While the 
model cubelets and summary plots may be useful for a variety of scientific applications, it is important to note that the key data products are the \wkapp\ model parameters and uncertainties from which the other data products are derived.


The \textsc{WKAPP} data products are accessible via both CSIRO ASKAP Science Data Archive (CASDA)\footnote{\href{https://research.csiro.au/casda/}{https://research.csiro.au/casda/}} and the Canadian Astronomy Data Centre (CADC)\footnote{\href{https://www.cadc-ccda.hia-iha.nrc-cnrc.gc.ca/}{https://www.cadc-ccda.hia-iha.nrc-cnrc.gc.ca/}}. A full description of the data access options can be found in both \citetalias{SoFiARelease} and through the data access portal\footnote{\href{https://wallaby-survey.org/data/}{https://wallaby-survey.org/data/}}.


\subsection{Model Limitations}
\label{subsec:mod_lim}

The procedure adopted here for producing kinematic models of the WALLABY PDR1 galaxies is a reasonable first effort.  However, it is important to note that there are limitations to both elements of the approach adopted as well as to the underlying data; we discuss them below, in what we judge to be decreasing order of importance from the perspective of using the \wkapp\ products.  Many of these issues will be solved in future releases through improved data analysis and a custom kinematic pipeline that is optimitized for WALLABY detections.

\begin{figure}
\centering
    \includegraphics[width=0.95\textwidth]{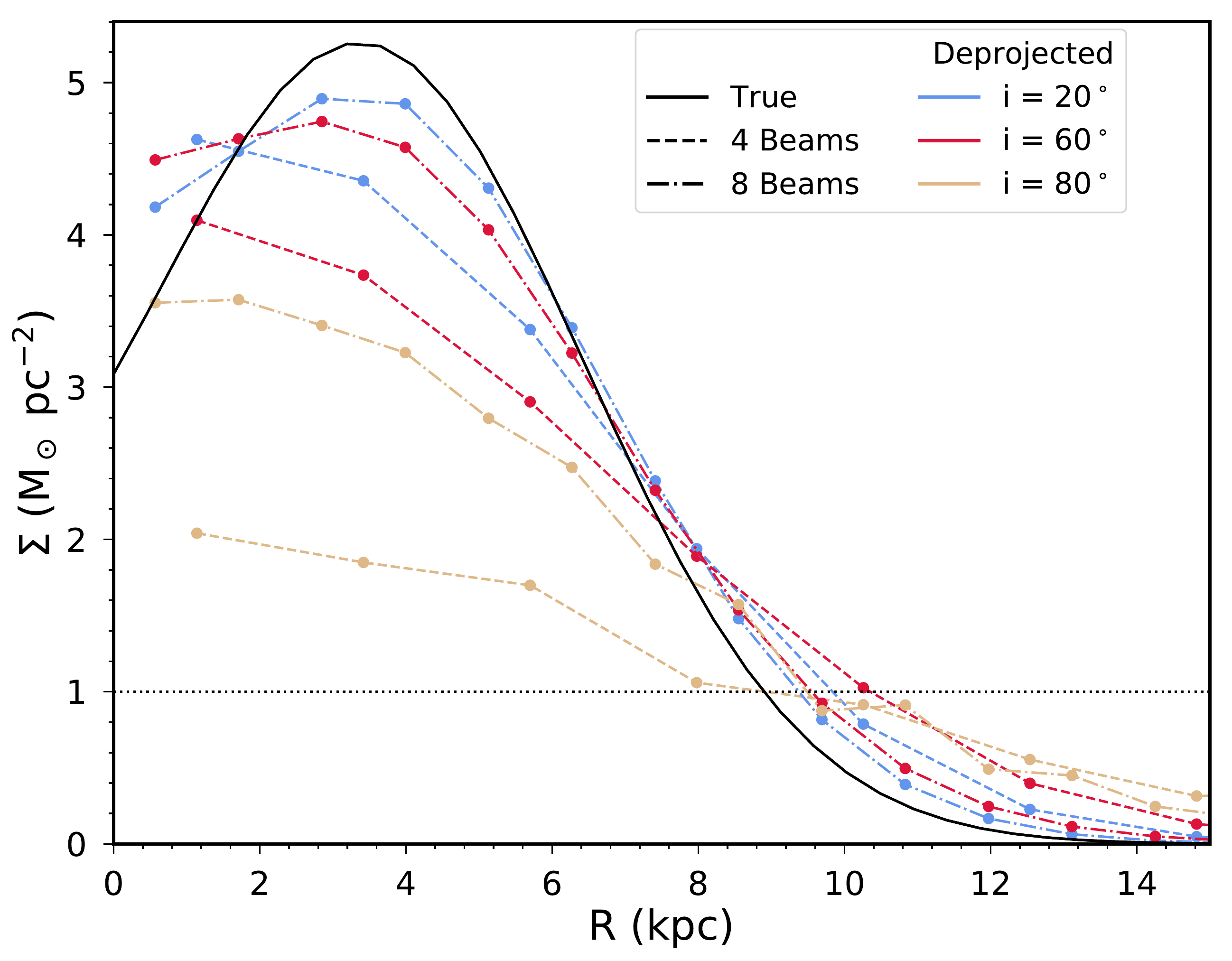} \caption{Comparison of the deprojected surface densities (dotted and dashed coloured lines) recovered from \textsc{WKAPP} of  a mock input surface density (solid black line) in PDR1-like sources that are resolved by $D_{HI}=4$ beams and $D_{HI}=8$  beams across their major axes at disk inclinations of $20^\circ$, $60^\circ$, and $80^\circ$.  }
  \label{Fig:SD_Test}
\end{figure}

The most obvious issue in the kinematic modelling approach is the deprojection of the surface density measured from the Moment 0 maps (see Sec.~\ref{subsec:mod_surfdens}).  The standard $\cos(i)$ correction adopted here is known to fail at high inclinations due to the line-of-sight passing through multiple radii (e.g. \citealt{Sancisi1979}).  The failure of the $\cos(i)$ correction in the PDR1 regime is illustrated very clearly in Fig. \ref{Fig:SD_Test}, which shows an input deprojected surface density distribution and the distributions that are recovered for PDR1-like sources generated using \mcgsuite\ at different resolutions and inclinations using the \textsc{WKAPP} method.

Fig. \ref{Fig:SD_Test} illustrates the well-known result that as the galaxy becomes more highly inclined and more poorly resolved, the deprojected surface density from ellipse fits to Moment 0 maps becomes much less reliable in both total flux and profile shape.  In particular, the inner profile peak is strongly underestimated as the inclination increases and the resolution decreases. We caution the user against using the inner deprojected surface density profile unless the impact of these biases is quantified for the particular application at hand.  

Conversely, the outer profile  profile shape is similar to the input one except at the poorest resolution and highest inclination shown in Fig. \ref{Fig:SD_Test}.  However, the profile extent is biased to larger $R$ due to beam smearing. We judge the outer deprojected surface densities and \hi\ sizes  to be robust enough for use in many cases, although \hi\ sizes should be first corrected for beam smearing \citep[e.g.][]{Wang2021,Reynolds2021}. In the future, the custom WALLABY pipeline will fit the surface density using the full 3D cube and so, should be more accurate than the ellipse fitting adopted here.

A second limitation is the restriction of the kinematic analysis to flat-disk models.  As described in Sec.~\ref{Sec:ModelCodeComp}, the homogeneous application of models to all suitable PDR1 detections is a key principle of our modelling approach, which drives the flat-disk modelling choice: in the marginally-resolved regime, it is often not possible to reliably explore warps, non-radial flows, and other complicated features due to a lack of statistically independent constraints on the underlying structure. Certainly, some of the more well-resolved galaxies in the sample show evidence for these complicated structures; for example, the slight offset in the minor axis PV diagram for J100426-282638 in Fig.~\ref{Fig:Summary} indicates some level of non-circular motions.  More sophisticated modelling of these objects is likely warranted, and well-suited to 2D analyses where non-axisymmetric structures can also be explored \citep[e.g.][]{Oh2018, Sellwood2021}. This work is underway for PDR1 sources. 

A related issue is that complicated structural features may be present but not spatially resolved, biasing the flat-disk models constructed here. The importance of this bias is not known at present, but can be constrained by convolving mock or real galaxies that exhibit such features down to the marginally-resolved regime and exploring how well flat-disk models recover their structure. While such tests have not yet been performed for PDR1 sources, they will be investigated in future data releases.

As a result of the second key principle that underpins our modelling approach and in light of the lack of statistical uncertainties returned by available tilted-ring algorithms (Sec.~\ref{Sec:ModelCodeComp}), the uncertainties on the optimized model are derived from the differences between the \barolo\ and \fat\ applications to the pre-processed PDR1 cubelets (Sec.~\ref{subsec:mod_application}). While we judge these uncertainties to be reasonable estimates of the reliability of the model parameters returned that can be used for scientific applications, they have not been vetted as statistical representations of the dispersion in model properties for the PDR1 galaxy population. As such, we recommend that they be considered as lower limits of the absolute uncertainty on the properties of the underlying \hi\ disks. The custom WALLABY pipeline that will be implemented in future data releases will include robustly determined statistical uncertainties through either monte-carlo or bootstrap-resampling approaches \citep[e.g.][]{Davis2013, Sellwood2021}.

Finally, we note that we have used the output PDR1 source cubelets from as inputs to WKAPP. \citetalias{SoFiARelease} discuss a number of data quality issues that may affect the PDR1 release (see their section~4). The most significant from a kinematic modelling perspective is likely the adoption of a $30''$ circular Gaussian beam even for sources that may not have been deconvolved during calibration and imaging. \textcolor{black}{This issue is likely to affect the poorly-resolved, lower-$S/N$ detections, which may be better characterized by the dirty beam, which has beam dimensions of $\sim 30" \times 27"$.}  Moreover, signatures of the dirty beam in the form of negative sidelobes may still remain in the cubelets.  Considering the small difference in dimensions between the dirty beam and the restored beam, the different beam sizes are unlikely to affect our kinematic models. It is more plausible that the presence of negative sidelobes from the dirty beam have biased the modelled disk morphologies, but since the first ASKAP sidelobe peaks at $\sim5\%$ of the main lobe response and since the integrated systematic effect on the measured fluxes is only of order $\sim20\%$ \citepalias{SoFiARelease}, we expect the effect on the disk morphologies to be mild. On balance, we conclude that both of these effects are likely to be insignificant relative to the other limitations in the kinematic models discussed above.

\section{Kinematic Model Catalog}
\label{Sec:CatalogueAndProducts}

We have successfully generated \wkapp\ kinematic models for 124 PDR1 sources; since 15/19 modelled Hydra TR1 sources also have models in Hydra TR2, WKAPP has produced kinematic models for a total of 109 unique PDR1 objects.  Table \ref{tab:Successes} lists the number of sources in each field and team release, the number of sources for which modelling was attempted, and the number for which successful models were obtained.  Considering that we attempted to model 209/592 ($35\%$) unique sources, our model success rate is $\sim 60\%$.  The coloured points in Fig.~\ref{Fig:Size-SNPlot} summarize these results in the source size-$\log(S/N)$ plane. The mean uncertainties on the geometric model parameters are listed in Table \ref{tab:Uncertainties}: we typically constrain the kinematic centre to a few arcsec and $\kms$, and the disk inclination and \textcolor{black}{position angle} to better than $\sim5^\circ$ and $\sim2^\circ$, respectively.

We note that the differences between the 15 unique objects in the Hydra field for which for both the TR1 and TR2 detections have successful kinematic models, the differences between them are generally small: the rotation curve differences are typically smaller than the uncertainties due to inclination. We recommend using the Hydra TR2 models over the Hydra TR1 models when both are available.  

\begin{table*}
    \centering
    \begin{tabular}{|c|c|c|c|c|c|}
        \hline
         & Hydra TR1 & Hydra TR2 & \textcolor{black}{Hydra Unique} & Norma TR1 & NGC 4636 TR1\\
         \hline 
        $N_{\rm{sources}}$ & 148 & 272 & 301 & 144 & 147 \\ 
        $N_{\rm{attempted}}$ & 37 & 74 & 79 & 63 & 67 \\
        $N_{\rm{success}}$ & 19 & 31 & 35 & 31 & 43\\
        \hline 
    \end{tabular}
    \caption{Number of PDR1 sources in each field (first row), the number for which WKAPP modelling was attempted (second row), and the number of successful WKAPP models (third row).}
    \label{tab:Successes}
\end{table*}


\begin{table}
    \centering
    \begin{tabular}{|c|c|}
        \hline
         Parameter & Mean Uncertainty \\
         \hline 
         \code{V\_sys} & $2.2~\kms$\\
         \code{Inc\_model} & $4.3^{\circ}$\\
         \code{PA\_model\_g} & $1.5^{\circ}$\\
         \code{RA\_model} & $2.4$\arcsec\\
         \code{DEC\_model} & $2.0$\arcsec\\
        \hline 
    \end{tabular}
    \caption{Mean uncertainties for the geometric parameters of optimized models. }
    \label{tab:Uncertainties}
\end{table}

\begin{figure*}
\centering
    \includegraphics[width=0.9\textwidth]{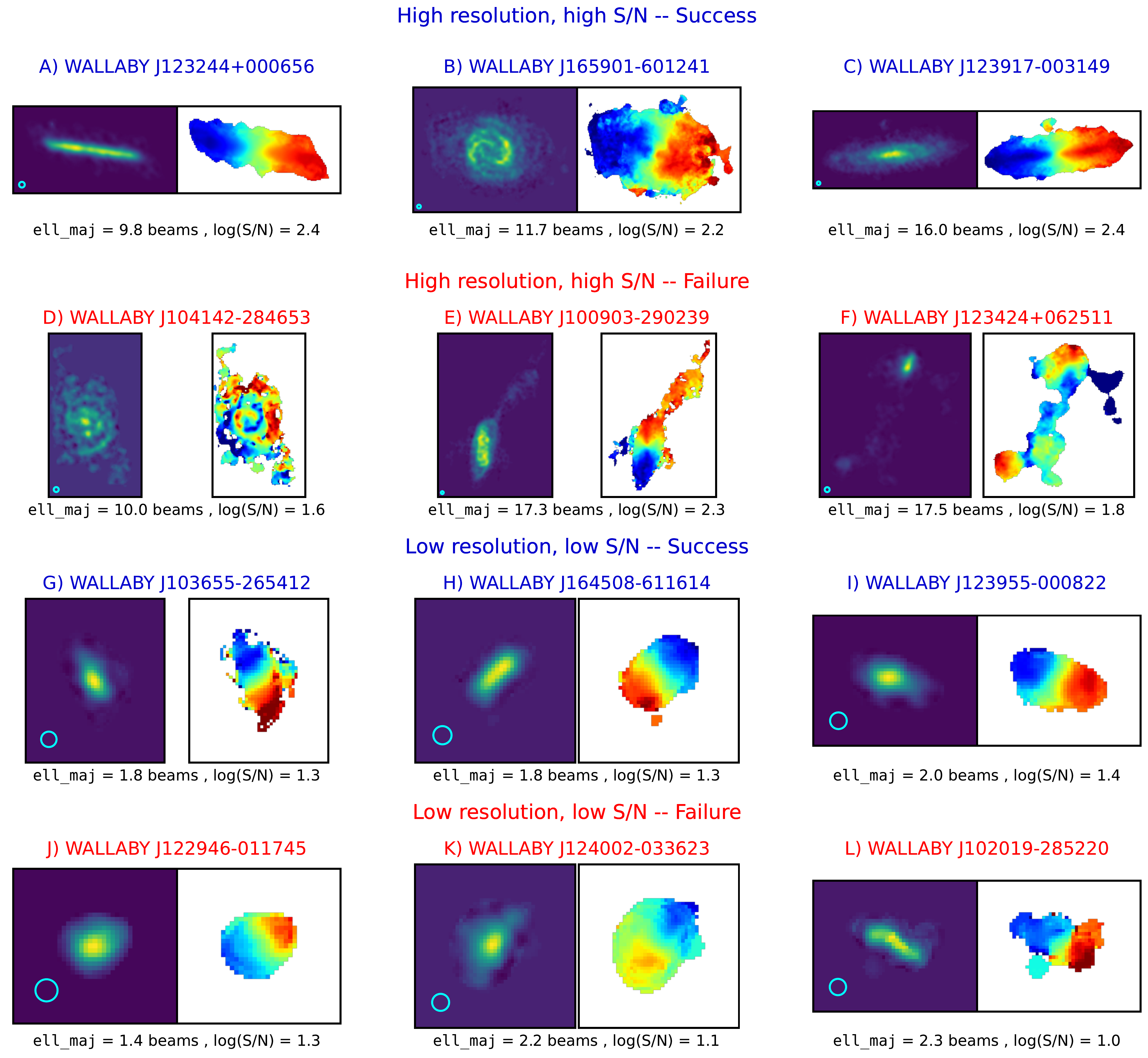} \caption{Moment 0 and Moment 1 maps for a sample set of PDR1 sources where the \textsc{WKAPP} modelling is a success (first and third rows) or a failure (second and fourth rows).  The top two rows show well-resolved and high-$S/N$ sources, while the bottom two rows have low resolution and $S/N$ values.  These sources are shown in Fig.~\ref{Fig:Size-SNPlot} with the outlined symbols.  The open ellipse in the Moment 0 maps shows the beam FWHM.  Figure \ref{Fig:Size-SNPlot} indicates each of the galaxies as open blue squares (top row), open red squares (second row), open blue diamonds (third row), and open red diamonds (bottom row).}
  \label{Fig:Montage}
\end{figure*}

To illustrate the conditions under which WKAPP succeeds and fails, Figure \ref{Fig:Montage} shows moment maps of PDR1 sources in both of these categories. These include both the highly resolved, high $S/N$ \textcolor{black}{(\code{ell\_maj} $\ge~7$ beams and $\log(S/N_{\mathrm{obs}})\ge 1.6$)} and the marginally resolved, low $S/N$ regimes \textcolor{black}{(\code{ell\_maj} $\le~2.5$ beams and $\log(S/N_{\mathrm{obs}})\le 1.4$)}.  
\textcolor{black}{The galaxies in the A-C panel pairs have morphologies consistent with rotating disks and do not show strong signatures of non-circular motions, asymmetries, disturbances, or other such features (although panel pair B does show spiral arms and small non-circular motions).  Given that our modelling method treats galaxies as `flat' disks, it is unsurprising that we successfully model this type of high $S/N$, high resolution detection.}
\textcolor{black}{Nonetheless,} each galaxy in the top row is a candidates for more detailed modelling in the future as they have sufficient resolution elements to identify warps, non-circular flows, measure the velocity dispersion, etc.  

The high resolution, high $S/N$ failures in Fig.~\ref{Fig:Montage} are all interesting, as each galaxy fails for a different reason.  \textcolor{black}{Panel pair D} shows a galaxy with a very complicated velocity profile that may be related to infalling \hi\ from a recent interaction.  \textcolor{black}{The E panel pair} shows a galaxy with an extended tidal tail; as explained in \citetalias{SoFiARelease}, this cubelet may also contain deconvolution artifacts.  This galaxy may be modelled in the future with slightly more careful masking/modelling.  \textcolor{black}{Panel pair F} show a pair of interacting galaxies that  \sofia\ detected as a single object.  As the masking improves \textcolor{black}{within \textsc{WKAPP}}, both those objects may be modelled in the future, \textcolor{black}{but such modelling will be challenging due to their interaction}. 

The low resolution, low $S/N$ rows are also quite interesting.  Unlike their higher resolution counterparts, it is more difficult to identify the reasons for the specific modelling successes or failures. \textcolor{black}{On the whole, however, we find that the most common cause of a failure in this regime is that the default source-finding mode in \fat\ or \barolo\ is unable to find the source in the cubelet.  Additionally, both \barolo\ and \fat\ have a number of default quality control flags, which a low resolution, low $S/N$ source may not satisfy.  Another situation where the codes may fail is when the automated initial \fat\ or \barolo\ estimate of the model parameters is poor, which then results in a poor fit.  It is again important to note here that both \fat\ and \barolo\ can be tuned individually to overcome these issues and produce accurate models for some of these cases, but that we elect to run both codes automatically and homogeneously on PDR1 sources (Sec.~\ref{sec:TRModelling})}.

Perhaps the most surprising result is the number of marginally resolved objects that have been modelled with only a few beams of resolution.  This is a testament to the power of 3D tilted ring modelling.  Contrasting the low resolution, low $S/N$ successes to the failures suggests that it is the combination of low $S/N$ \textit{and} low resolution that leads to the modelling efforts failing for apparently similar objects (based on a visual comparison of the moment maps). This suggests that the size and $S/N$ cuts applied to the PDR1 detection sample will be sufficient for future modelling efforts.

\begin{figure*}
\centering
    \includegraphics[width=0.65\textwidth]{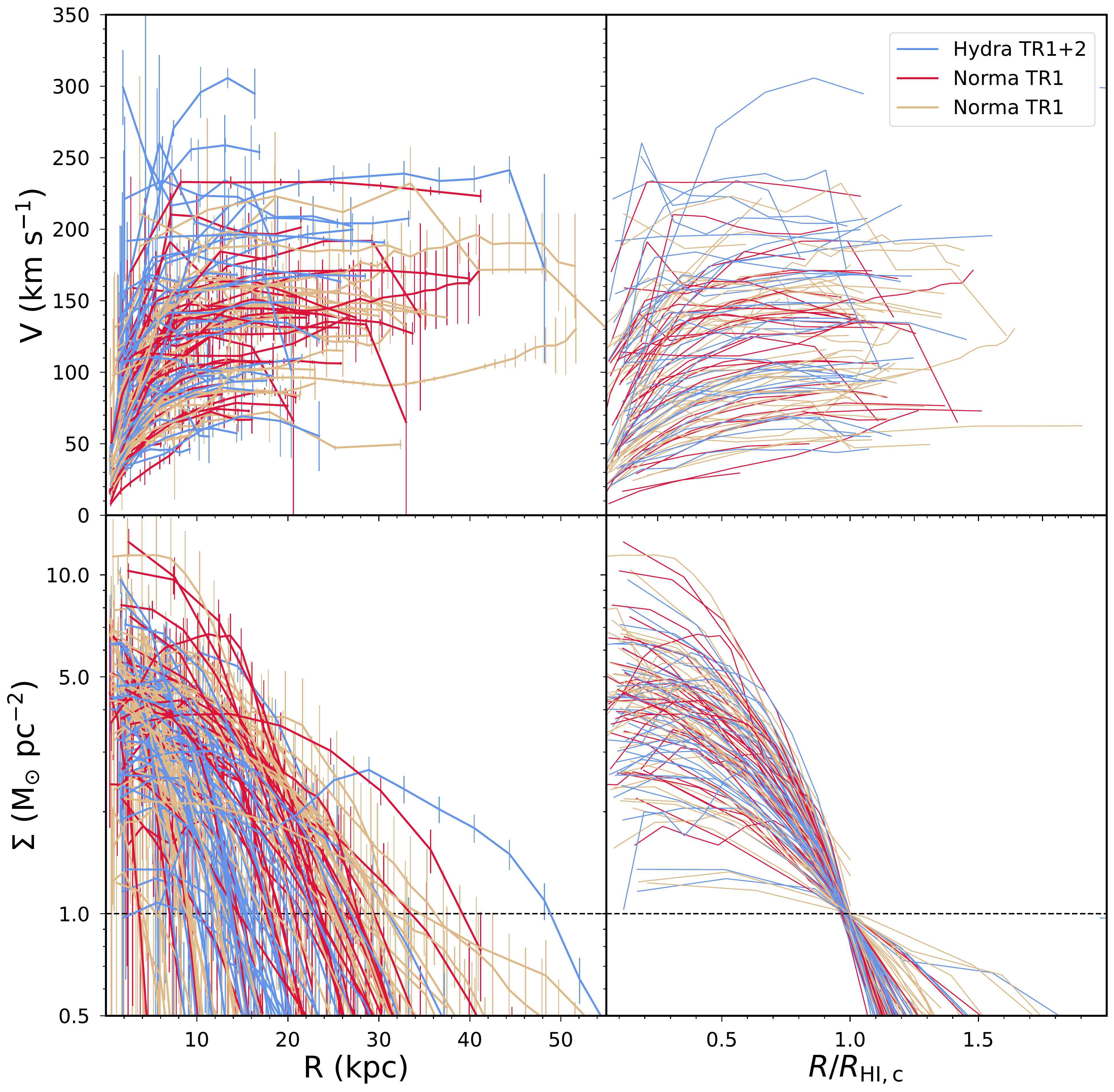} \caption{The full sample of optimized model rotation curves (top panel) and deprojected surface density profiles (bottom panel) for all pilot fields.  The blue, red, and orange lines show galaxies from the Hydra, Norma, and NGC 4636 fields respectively.  The radial sizes are calculated using redshift derived distances.  The horizontal dashed line shows the $1 $M$_{\odot}$ pc$^{-2}$ surface density that defines the \hi\ radius of a galaxy. The right-hand panels are normalized by $R_{\hi,c}$ as determined from the surface density profiles.}
  \label{Fig:RCs}
\end{figure*}

Figure \ref{Fig:RCs} illustrates the diversity of the modelled galaxy rotation curves and surface density profiles across PDR1 detections. When source distances are calculated from barycentric redshifts under the assumption of a flat $\Lambda$CDM cosmology with a Hubble parameter of $H_{0} = 70~\mathrm{km \, s^{-1} \, Mpc^{-1}}$ and a local matter density of $\Omega_{\rm m} = 0.3$, the measured sizes range from a few to tens of kiloparsecs.  The rotation velocity amplitudes range from $30-250~\kms$. Such a range in size, velocity, and from inference, mass, means that this sample of kinematic models will be valuable for many different studies and science questions.  \textcolor{black}{This sample is of comparable size and covers a similar mass range as SPARC (Spitzer Photometry and Accurate Rotation Curves; \citealt{Lelli2016}), which contains 175 rotation curves. SPARC and the PDR1 kinematic release are highly complementary across a range of scientific applications: while SPARC galaxies are generally better resolved than PDR1 sources, the PDR1 selection function is well-defined \citepalias{SoFiARelease}. }

\textcolor{black}{The deprojected surface densities in the lower panels of Fig. \ref{Fig:RCs}  suggest that there is an \hi\ surface density saturation at $\sim 10 \textrm{M}_{\odot}~\textrm{pc}^{-2}$, consistent with the results of \citet{Bigiel2008}}.  However, we  caution against using the inner surface densities for scientific applications without further modelling, given the breakdown in the standard $\cos(i)$ correction used to derive the deprojected surface densities in the marginally-resolved regime (see Section \ref{subsec:mod_lim} and Fig.~\ref{Fig:SD_Test}).  

\begin{figure}
\centering
    \includegraphics[width=0.8\textwidth]{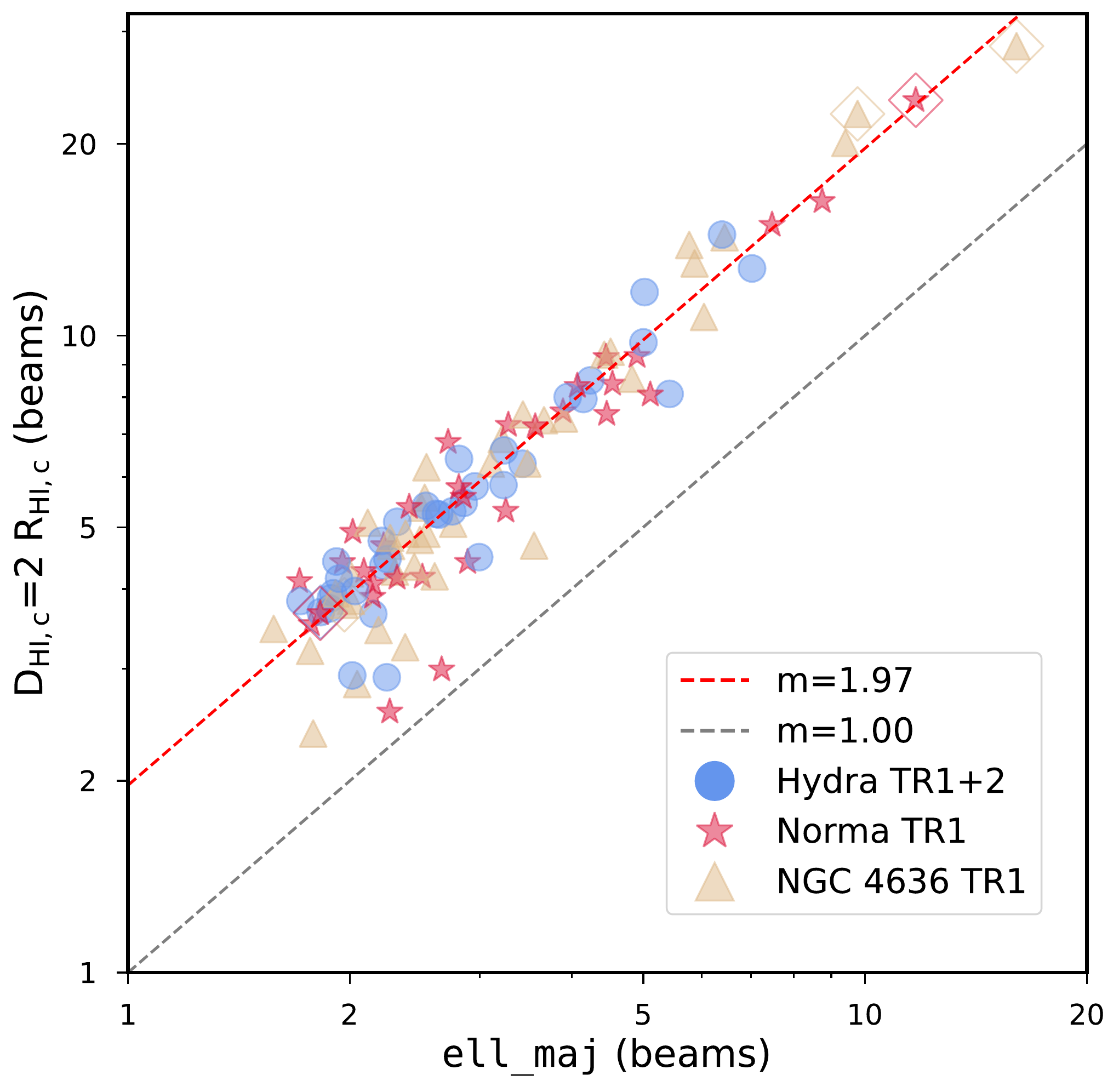} \caption{A comparison of the $R_{\hi}$ radius to the \textsc{SoFiA} \ellmaj\ parameter for all successfully modelled PDR1 detections.  The black dashed line shows the one-to-one line, the red dashed line shows the best fit straight line to the data, while the circle, star, and triangle symbols indicate galaxies in the Hydra, Norma, and NGC 4636 fields respectively. \textcolor{black}{The values for $m$ are the slopes of the one-to-one and best fit lines in linear space, respectively.} The open symbols indicate the fitted galaxies (rows 1 and 3) shown in Fig. \ref{Fig:Montage}.}
  \label{Fig:Ell_Maj_RHI}
\end{figure}

\textcolor{black}{By contrast, the outer deprojected surface density profiles are reliably recovered by WKAPP, modulo being radially smeared by the beam (see Section \ref{subsec:mod_lim} and Fig.~\ref{Fig:SD_Test}). The outer profiles in Fig.~\ref{Fig:RCs} show the characteristic exponential outer decline noted in previous work \citep[e.g.][]{Wang16}. We plot in Figure~\ref{Fig:Ell_Maj_RHI} the  diameter $D_{HI,c}$ at which the deprojected surface crosses $1\,\mathrm{M_\odot\,pc^{-2}}$ as a function of \ellmaj\ recovered by \sofia\ (see Fig.~\ref{Fig:Size-SNPlot}).} We note that since both parameters are estimates of disk size from the PDR1 Moment 0 maps, they should be similarly beam smeared. 

\textcolor{black}{The best fitting line for the data (performed in linear space) of $m=1.97$ shown in Fig.~\ref{Fig:Ell_Maj_RHI} illustrates that, in general,} $D_{HI,c}$ exceeds \ellmaj\ by a factor of $\sim$two. \ellmaj\ is computed from the second spatial moment of the Moment 0 map along the major axis \citep{Serra2015}, which for a Gaussian profile approaches twice its standard deviation. The factor of $\sim$2 difference between $D_{HI,c}$ and \ellmaj\ then arises naturally from the outer Gaussian profile shape provided it peaks in the range $6\,\mathrm{M_\odot\,pc^{-2}} - 10\,\mathrm{M_\odot\,pc^{-2}}$, which is generally the case for the PDR1 sources plotted in Fig.~\ref{Fig:RCs}. \textcolor{black}{This difference between \ellmaj\ and $D_{\hi}$ justifies our PDR1 selection criterion of \ellmaj\ $\,>\,2$ beams (see Sec.~\ref{subsec:gal_selection}). Figs.~\ref{Fig:Ell_Maj_RHI}~and~\ref{Fig:Size-SNPlot} imply that the majority of successful kinematic models have $D_{\hi} \ge 3.5$ beams, consistent with the modelling tests of \citet{Lewis2019}.}



\section{The Population of Kinematically Modelled Detections}
\label{Sec:Populations}

A key question to ask when producing a survey is what are the biases in a particular sample?  In this case, are there any biases/selection effects that apply to the kinematically modelled sample of galaxies relative to the larger WALLABY sample?  

\begin{figure*}
\centering
    \includegraphics[width=0.8\textwidth]{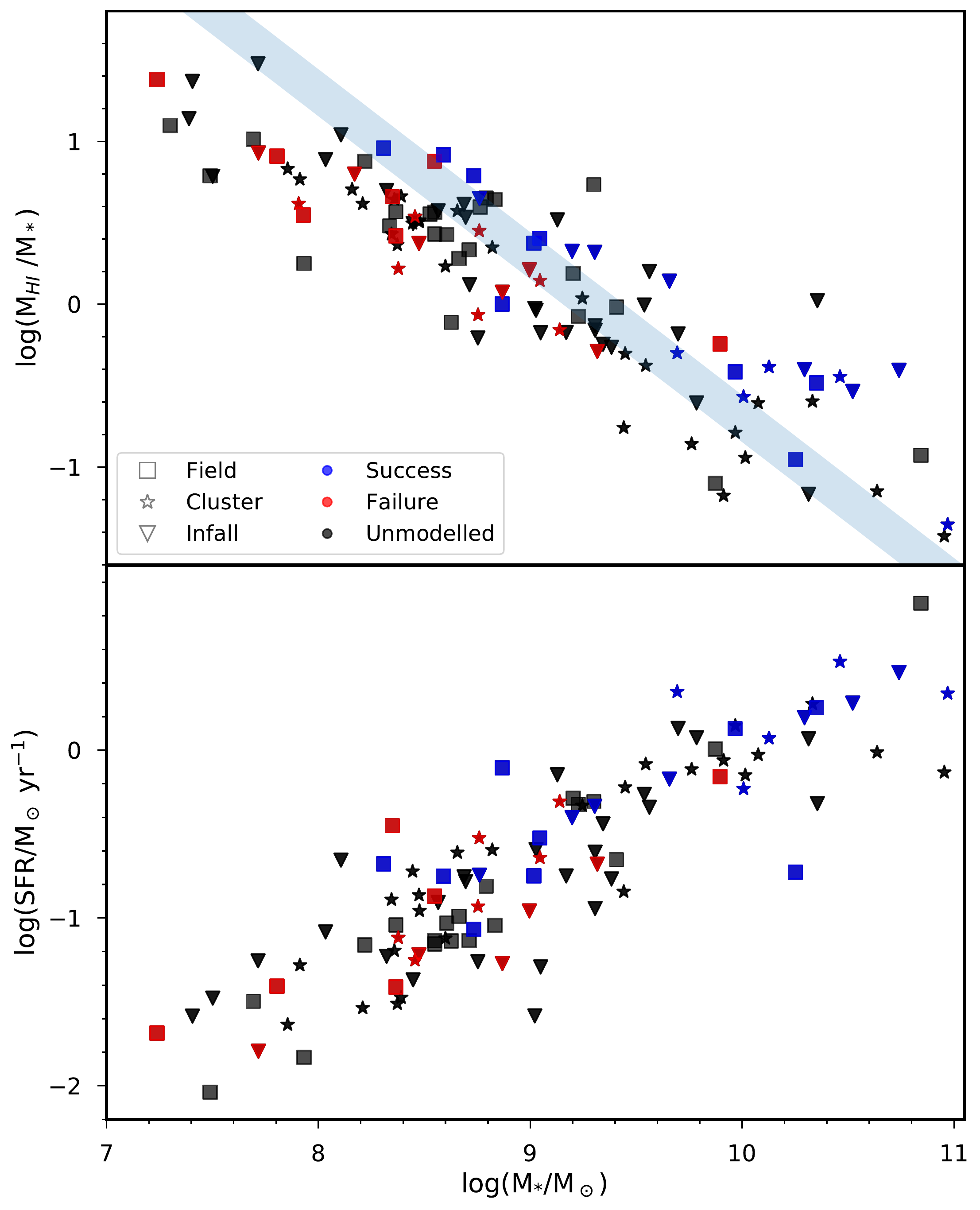} \caption{The population of kinematically modelled PDR1 detections in the Hydra field in the context of the $\hi$ mass - stellar mass relation (top panel) and star-formation - stellar mass relation (bottom panel) from \citep{Reynolds2022}.  In both panels, the symbol \textcolor{black}{shape} denotes the environmental designation from \citet{Reynolds2022}. Galaxies that have been successfully kinematically modelled are plotted in blue, while those for which modelling was attempted but failed are plotted in red.  In the top panel, the  blue line shows the predicted locus of points for galaxies that lie on the \hi-mass - \hi-diameter relation \citep{Wang16} with \hi\ diameters between 3 and 4 beams across at the at the Hydra cluster distance ($D=60\,\mathrm{Mpc}$, \citealt{Jorgensen1996}). The galaxies that were successfully modelled tend to lie above that region, indicating that one of the main drivers of modellability is angular size. There is no qualitative correlation between environment, SFR and galaxy modellability in the sample examined.}
  \label{Fig:Population}
\end{figure*}

To investigate the possibility of an environmental selection, we used the \citet{Reynolds2022} dataset of Hydra TR2 WALLABY detections  with velocities $cz<7000$ km/s.  In that work, \citet{Reynolds2022} classifed the galaxy environment as field, infalling, or cluster.  The top panel of Figure \ref{Fig:Population}  shows these galaxies along with their measured stellar and $\hi$ masses, while the bottom panel shows successful and failed models in the star formation rate -- stellar mass plane.  We find no qualitative evidence that galaxies in different environments are more or less likely to be modellable, though the sample is relatively small to subdivide by environment.  Morevoer, such environmental effects are difficult to discern using only detections in HI-blind, shallow surveys due to selection effects \citep{Cortese2021}.

However, Fig. \ref{Fig:Population} suggests that kinematic models of sources with $M_\star \le 10^{8.5}\,M_\odot$ tend to fail, models of sources with $M_\star \ge 10^{9.5}\,M_\odot$ tend to succeed, and successfully modelled sources with $M_{\star} \sim 10^9\,M_\odot$ tend to be more gas-rich than sources where the models failed. These trends are a consequence of the \hi\ mass - \hi\ size relation \citep{Wang16}, the locus of which for Hydra cluster galaxies spatially resolved by 3--4 beams is shown by the shaded region in the top panel. Model successes tend to lie above the shaded region, while failures tend to lie below it. This threshold is broadly consistent with the benchmark results of \citet{Lewis2019} for \fat, and demonstrates that galaxy angular sizes (and $S/N$ by virtue of the correlation in Fig. \ref{Fig:Size-SNPlot}) are the strongest predictors of modellability with WKAPP among PDR1 sources. 


\section{Conclusions}
\label{Sec:Conclusions}

WALLABY will use the ASKAP telescope to detect the \hi\ content of $\sim 210,000$ galaxies out to redshift $z \sim 0.1$.  The PDR1 observations target three fields observed at the full resolution and sensitivity of the planned survey.  The source-finding analysis applied to these fields detected 592  \hi\ sources \citepalias{SoFiARelease}.  Of those, we have kinematically modelled 109 galaxies.  

In our modelling approach (\textsc{WKAPP}), we attempt to fit all detections with $\ellmaj \ge 2$ beams or $\log(S/N_{obs}\ge 1.25$ using both \barolo\ and \fat.  There are 209 unique galaxies that meet this criteria.  Both the \barolo\ and \fat\ analyses are constrained to consider only pure flat disk models in order to obtain a uniform and robust population of modelled galaxies.  The results of the individual applications are examined visually to determine their plausibility.

The optimized PDR1 models are generated by first averaging the geometric parameters derived from the \fat\ and \barolo\ fits.  Then the inclination corrected, interpolated rotation curves are averaged together to generate the optimized rotation curve.  Finally, the surface density is extracted from the \sofia\ masked Moment 0 map via ellipse fitting using the optimized model geometry.

The full set of kinematic models are publicly available at \textcolor{black}{the WALLABY pilot phase data portal}.  The modelled population tends to be gas-rich and tends to have larger stellar masses than the non-modelled population.  This is largely expected from the \hi\ mass-size relation.

The \textsc{WKAPP} modelling success rate is roughly $20\%$ (109/592). This success bodes well for the full WALLABY dataset.  The $\sim20\%$ modelling suggests that we will generate kinematic models for $\sim 40,000$ galaxies over the full survey.  However, the three PDR1 fields were chosen for testing purposes and contain galaxies that may be \textcolor{black}{less distant} than the average WALLABY galaxy.  Given the importance of size and $S/N$ the full WALLABY modelling success rate may be somewhat lower than $20\%$, but it is still likely higher than $10\%$ \citep{Koribalski2020}.

The use of \textsc{WKAPP} on the PDR1 sources has been quite successful.  \textcolor{black}{While the modelling success rate across the full sample is only $\sim$20\%, for sources with \ellmaj$\ge 2$ beams it is $\sim 50\%$.  Additionally, galaxies with \ellmaj$\ge 2$ beams are less resolved than prior estimates of the \fat\ and \barolo\ resolution limits, allowing us to attempt kinematic models on a greater number of sources than initially expected.  Beyond the successes of \textsc{WKAPP} for PDR1 sample, }it is a critical step in developing the full, automatic pipeline that will be deployed for the full WALLABY survey.  In the meantime, these kinematic models are useful for a large variety of science investigations.  Moreover, examining the population of galaxies where the models failed is also informative, and has revealed many intriguing and complicated objects.  While the kinematic models presented here are, by necessity, relatively simple, there are a number of candidates for more detailed \textcolor{black}{2D and 3D} modelling. \textcolor{black}{Comparing the \textsc{WKAPP} models to existing models for some of these candidates as well as exploring more detailed 2D and 3D modelling efforts will help to understand the strengths and weaknesses of this approach.  Another important exercise will be testing \textsc{WKAPP}, the future full pipeline, and other kinematic modelling software packages, using mock WALLABY observations from large cosmological simulations.}

\textcolor{black}{The PDR1 kinematic models presented here are the first step towards full set of WALLABY kinematic models.  We plan to publically release the rotation curves, surface density profiles, and other properties for the $10000-40000$ galaxies that we expect to model.  This will form a large, homogeneous legacy data set that will allow explorations of the velocity function, the TF relation, investigations of galaxy mass distributions, and much more.  }

\begin{acknowledgement}

\textcolor{black}{We would like to thank the referee for their useful comments and suggestions for the paper.}

   \textcolor{black}{The Australian SKA Pathfinder is part of the Australia Telescope National Facility (\href{https://ror.org/05qajvd42}{https://ror.org/05qajvd42}) which is managed by CSIRO. Operation of ASKAP is funded by the Australian Government with support from the National Collaborative Research Infrastructure Strategy. ASKAP uses the resources of the Pawsey Supercomputing Centre. Establishment of ASKAP, the Murchison Radio-astronomy Observatory and the Pawsey Supercomputing Centre are initiatives of the Australian Government, with support from the Government of Western Australia and the Science and Industry Endowment Fund. We acknowledge the Wajarri Yamatji as the traditional owners of the Observatory site.}
    
    \textcolor{black}{WALLABY acknowledges technical support from the Australian SKA Regional Centre (AusSRC) and Astronomy Data And Computing Services (ADACS).}
    
    This research used the facilities of the Canadian Astronomy Data Centre operated by the National Research Council of Canada with the support of the Canadian Space Agency.
    
This paper includes archived data obtained through the CSIRO ASKAP Science Data Archive, CASDA (\href{http://data. csiro.au}{http://data. csiro.au}).

 This paper uses resources from the  Canadian Initiative for Radio Astronomy Data Analysis (CIRADA), which is funded by a grant from the Canada Foundation for Innovation 2017 Innovation Fund (Project 35999) and by the Provinces of Ontario, British Columbia, Alberta, Manitoba and Quebec, in collaboration with the National Research Council of Canada, the US National Radio Astronomy Observatory and Australia’s Commonwealth Scientific and Industrial Research Organisation.

Part of this research was conducted by the Australian Research Council Centre of Excellence for All Sky Astrophysics in 3 Dimensions (ASTRO 3D), through project number CE170100013.

AB acknowledges support from the Centre National d’Etudes Spatiales (CNES), France. EDT was supported by the US National Science Foundation under grant 1616177.  JMvdH acknowledges support from the European Research Council under the European Union’s Seventh Framework Programme (FP/2007-2013) / ERC Grant Agreement nr. 291531 (HIStoryNU).  KS acknowledges support from the Natural Sciences and Engineering Research Council of Canada (NSERC).  LVM acknowledges financial support from the State Agency for Research of the Spanish Ministry of Science, Innovation and Universities through the "Center of Excellence Severo Ochoa" awarded to the Instituto de Astrofísica de Andalucía (SEV-2017-0709), from grant RTI2018-096228-B-C31 (MCIU/AEI/FEDER,UE), from the grant IAA4SKA (Ref. R18-RT-3082) from the Economic Transformation, Industry, Knowledge and Universities Council of the Regional Government of Andalusia and the European Regional Development Fund from the European Union.  PK is partially supported by the BMBF project 05A17PC2 for D-MeerKAT. SHOH acknowledges a support from the National Research Foundation of Korea (NRF) grant funded by the Korea government (Ministry of Science and ICT: MSIT) (No. NRF-2020R1A2C1008706).  TS acknowledges support by Funda\c{c}\~{a}o para a Ci\^{e}ncia e a Tecnologia (FCT) through national funds (UID/FIS/04434/2013), FCT/MCTES through national funds (PIDDAC) by this grant UID/FIS/04434/2019 and by FEDER through COMPETE2020 (POCI-01-0145-FEDER-007672). TS also acknowledges the support by the fellowship SFRH/BPD/103385/2014 funded by the FCT (Portugal) and POPH/FSE (EC). TS additionally acknowledges support from DL 57/2016/CP1364/CT0009 from The Centro de Astrof\'{i}sica da Universidade do Porto.

This research uses  Astropy,\footnote{\href{http://www.astropy.org}{http://www.astropy.org}} a community-developed core Python package for Astronomy \citep{astropy:2013, astropy:2018}.  It also uses the Numpy \citep{Numpy}, SciPy \citep{SciPy}, PANDAS \citep{PANDAS}, and MatPlotLib \citep{Matplotlib} libraries.

\end{acknowledgement}


\bibliography{KinematicModels}

\begin{thebibliography}{}
\expandafter\ifx\csname natexlab\endcsname\relax\def\natexlab#1{#1}\fi

\bibitem[{{Astropy Collaboration} {et~al.}(2013){Astropy Collaboration},
  {Robitaille}, {Tollerud}, {Greenfield}, {Droettboom}, {Bray}, {Aldcroft},
  {Davis}, {Ginsburg}, {Price-Whelan}, {Kerzendorf}, {Conley}, {Crighton},
  {Barbary}, {Muna}, {Ferguson}, {Grollier}, {Parikh}, {Nair}, {Unther},
  {Deil}, {Woillez}, {Conseil}, {Kramer}, {Turner}, {Singer}, {Fox}, {Weaver},
  {Zabalza}, {Edwards}, {Azalee Bostroem}, {Burke}, {Casey}, {Crawford},
  {Dencheva}, {Ely}, {Jenness}, {Labrie}, {Lim}, {Pierfederici}, {Pontzen},
  {Ptak}, {Refsdal}, {Servillat}, \& {Streicher}}]{astropy:2013}
{Astropy Collaboration}, {Robitaille}, T.~P., {Tollerud}, E.~J., {et~al.} 2013,
  \aap, 558, A33

\bibitem[{{Astropy Collaboration} {et~al.}(2018){Astropy Collaboration},
  {Price-Whelan}, {Sip{\H{o}}cz}, {G{\"u}nther}, {Lim}, {Crawford}, {Conseil},
  {Shupe}, {Craig}, {Dencheva}, {Ginsburg}, {VanderPlas}, {Bradley},
  {P{\'e}rez-Su{\'a}rez}, {de Val-Borro}, {Aldcroft}, {Cruz}, {Robitaille},
  {Tollerud}, {Ardelean}, {Babej}, {Bach}, {Bachetti}, {Bakanov}, {Bamford},
  {Barentsen}, {Barmby}, {Baumbach}, {Berry}, {Biscani}, {Boquien}, {Bostroem},
  {Bouma}, {Brammer}, {Bray}, {Breytenbach}, {Buddelmeijer}, {Burke},
  {Calderone}, {Cano Rodr{\'\i}guez}, {Cara}, {Cardoso}, {Cheedella}, {Copin},
  {Corrales}, {Crichton}, {D'Avella}, {Deil}, {Depagne}, {Dietrich}, {Donath},
  {Droettboom}, {Earl}, {Erben}, {Fabbro}, {Ferreira}, {Finethy}, {Fox},
  {Garrison}, {Gibbons}, {Goldstein}, {Gommers}, {Greco}, {Greenfield},
  {Groener}, {Grollier}, {Hagen}, {Hirst}, {Homeier}, {Horton}, {Hosseinzadeh},
  {Hu}, {Hunkeler}, {Ivezi{\'c}}, {Jain}, {Jenness}, {Kanarek}, {Kendrew},
  {Kern}, {Kerzendorf}, {Khvalko}, {King}, {Kirkby}, {Kulkarni}, {Kumar},
  {Lee}, {Lenz}, {Littlefair}, {Ma}, {Macleod}, {Mastropietro}, {McCully},
  {Montagnac}, {Morris}, {Mueller}, {Mumford}, {Muna}, {Murphy}, {Nelson},
  {Nguyen}, {Ninan}, {N{\"o}the}, {Ogaz}, {Oh}, {Parejko}, {Parley}, {Pascual},
  {Patil}, {Patil}, {Plunkett}, {Prochaska}, {Rastogi}, {Reddy Janga},
  {Sabater}, {Sakurikar}, {Seifert}, {Sherbert}, {Sherwood-Taylor}, {Shih},
  {Sick}, {Silbiger}, {Singanamalla}, {Singer}, {Sladen}, {Sooley},
  {Sornarajah}, {Streicher}, {Teuben}, {Thomas}, {Tremblay}, {Turner},
  {Terr{\'o}n}, {van Kerkwijk}, {de la Vega}, {Watkins}, {Weaver}, {Whitmore},
  {Woillez}, {Zabalza}, \& {Astropy Contributors}}]{astropy:2018}
{Astropy Collaboration}, {Price-Whelan}, A.~M., {Sip{\H{o}}cz}, B.~M., {et~al.}
  2018, \aj, 156, 123

\bibitem[{{Battaner} {et~al.}(1990){Battaner}, {Florido}, \&
  {Sanchez-Saavedra}}]{Battaner1990}
{Battaner}, E., {Florido}, E., \& {Sanchez-Saavedra}, M.~L. 1990, \aap, 236, 1

\bibitem[{{Begeman}(1987)}]{Begeman1987}
{Begeman}, K.~G. 1987, PhD thesis, University of Groningen, Netherlands

\bibitem[{{Bekiaris} {et~al.}(2016){Bekiaris}, {Glazebrook}, {Fluke}, \&
  {Abraham}}]{Bekiaris2016}
{Bekiaris}, G., {Glazebrook}, K., {Fluke}, C.~J., \& {Abraham}, R. 2016,
  \mnras, 455, 754

\bibitem[{{Bigiel} {et~al.}(2008){Bigiel}, {Leroy}, {Walter}, {Brinks}, {de
  Blok}, {Madore}, \& {Thornley}}]{Bigiel2008}
{Bigiel}, F., {Leroy}, A., {Walter}, F., {et~al.} 2008, \aj, 136, 2846

\bibitem[{{Bosma}(1978)}]{bosma78}
{Bosma}, A. 1978, PhD thesis, University of Groningen, Netherlands

\bibitem[{{Cortese} {et~al.}(2021){Cortese}, {Catinella}, \&
  {Smith}}]{Cortese2021}
{Cortese}, L., {Catinella}, B., \& {Smith}, R. 2021, \pasa, 38, e035

\bibitem[{{Davis} {et~al.}(2013){Davis}, {Alatalo}, {Bureau}, {Cappellari},
  {Scott}, {Young}, {Blitz}, {Crocker}, {Bayet}, {Bois}, {Bournaud}, {Davies},
  {de Zeeuw}, {Duc}, {Emsellem}, {Khochfar}, {Krajnovi{\'c}}, {Kuntschner},
  {Lablanche}, {McDermid}, {Morganti}, {Naab}, {Oosterloo}, {Sarzi}, {Serra},
  \& {Weijmans}}]{Davis2013}
{Davis}, T.~A., {Alatalo}, K., {Bureau}, M., {et~al.} 2013, \mnras, 429, 534

\bibitem[{{de Blok}(2010)}]{deBlok2010}
{de Blok}, W.~J.~G. 2010, Advances in Astronomy, 2010, 789293

\bibitem[{{Di Teodoro} \& {Fraternali}(2015)}]{diTeodoro15}
{Di Teodoro}, E.~M., \& {Fraternali}, F. 2015, \mnras, 451, 3021

\bibitem[{{Di Teodoro} \& {Peek}(2021)}]{DiTeoDoro2021}
{Di Teodoro}, E.~M., \& {Peek}, J.~E.~G. 2021, \apj, 923, 220

\bibitem[{Harris {et~al.}(2020)Harris, Millman, van~der Walt, Gommers,
  Virtanen, Cournapeau, Wieser, Taylor, Berg, Smith, Kern, Picus, Hoyer, van
  Kerkwijk, Brett, Haldane, del R{\'{i}}o, Wiebe, Peterson,
  G{\'{e}}rard-Marchant, Sheppard, Reddy, Weckesser, Abbasi, Gohlke, \&
  Oliphant}]{Numpy}
Harris, C.~R., Millman, K.~J., van~der Walt, S.~J., {et~al.} 2020, Nature, 585,
  357

\bibitem[{{Hotan} {et~al.}(2021){Hotan}, {Bunton}, {Chippendale}, {Whiting},
  {Tuthill}, {Moss}, {McConnell}, {Amy}, {Huynh}, {Allison}, {Anderson},
  {Bannister}, {Bastholm}, {Beresford}, {Bock}, {Bolton}, {Chapman}, {Chow},
  {Collier}, {Cooray}, {Cornwell}, {Diamond}, {Edwards}, {Feain}, {Franzen},
  {George}, {Gupta}, {Hampson}, {Harvey-Smith}, {Hayman}, {Heywood}, {Jacka},
  {Jackson}, {Jackson}, {Jeganathan}, {Johnston}, {Kesteven}, {Kleiner},
  {Koribalski}, {Lee-Waddell}, {Lenc}, {Lensson}, {Mackay}, {Mahony},
  {McClure-Griffiths}, {McConigley}, {Mirtschin}, {Ng}, {Norris}, {Pearce},
  {Phillips}, {Pilawa}, {Raja}, {Reynolds}, {Roberts}, {Roxby}, {Sadler},
  {Shields}, {Schinckel}, {Serra}, {Shaw}, {Sweetnam}, {Troup}, {Tzioumis},
  {Voronkov}, \& {Westmeier}}]{Hotan2021}
{Hotan}, A.~W., {Bunton}, J.~D., {Chippendale}, A.~P., {et~al.} 2021, \pasa,
  38, e009

\bibitem[{Hunter(2007)}]{Matplotlib}
Hunter, J.~D. 2007, Computing in Science \& Engineering, 9, 90

\bibitem[{{Jones} {et~al.}(2021){Jones}, {Vergani}, {Romano}, {Ginolfi},
  {Fudamoto}, {B{\'e}thermin}, {Fujimoto}, {Lemaux}, {Morselli}, {Capak},
  {Cassata}, {Faisst}, {Le F{\`e}vre}, {Schaerer}, {Silverman}, {Yan},
  {Boquien}, {Cimatti}, {Dessauges-Zavadsky}, {Ibar}, {Maiolino}, {Rizzo},
  {Talia}, \& {Zamorani}}]{Jones2021}
{Jones}, G.~C., {Vergani}, D., {Romano}, M., {et~al.} 2021, \mnras, 507, 3540

\bibitem[{{J{\o}rgensen} {et~al.}(1996){J{\o}rgensen}, {Franx}, \&
  {Kj{\ae}rgaard}}]{Jorgensen1996}
{J{\o}rgensen}, I., {Franx}, M., \& {Kj{\ae}rgaard}, P. 1996, \mnras, 280, 167

\bibitem[{{J{\'o}zsa} {et~al.}(2007){J{\'o}zsa}, {Kenn}, {Klein}, \&
  {Oosterloo}}]{Jorza2007}
{J{\'o}zsa}, G.~I.~G., {Kenn}, F., {Klein}, U., \& {Oosterloo}, T.~A. 2007,
  \aap, 468, 731

\bibitem[{{J{\'o}zsa} {et~al.}(2009){J{\'o}zsa}, {Oosterloo}, {Morganti},
  {Klein}, \& {Erben}}]{Jorza2009}
{J{\'o}zsa}, G.~I.~G., {Oosterloo}, T.~A., {Morganti}, R., {Klein}, U., \&
  {Erben}, T. 2009, \aap, 494, 489

\bibitem[{{J{\'o}zsa} {et~al.}(2021){J{\'o}zsa}, {Thorat}, {Kamphuis},
  {Sebokolodi}, {Maina}, {Wang}, {Pieterse}, {Groot}, {Ramaila}, {Serra},
  {Andati}, {de Blok}, {Hugo}, {Kleiner}, {Maccagni}, {Makhathini},
  {Moln{\'a}r}, {Ramatsoku}, {Smirnov}, {Bloemen}, {Paterson}, {Vreeswijk},
  {McBride}, {Klein-Wolt}, {Woudt}, {K{\"o}rding}, {Le Poole}, {Goedhart},
  {Passmoor}, {Serylak}, \& {Dettmar}}]{Jorzsa2021}
{J{\'o}zsa}, G. I.~G., {Thorat}, K., {Kamphuis}, P., {et~al.} 2021, \mnras,
  501, 2704

\bibitem[{{Kamphuis} {et~al.}(2015){Kamphuis}, {J{\'o}zsa}, {Oh}, {Spekkens},
  {Urbancic}, {Serra}, {Koribalski}, \& {Dettmar}}]{Kamphuis2015}
{Kamphuis}, P., {J{\'o}zsa}, G.~I.~G., {Oh}, S. .~H., {et~al.} 2015, \mnras,
  452, 3139

\bibitem[{{Khoperskov} {et~al.}(2014){Khoperskov}, {Moiseev}, {Khoperskov}, \&
  {Saburova}}]{Khoperskov2014}
{Khoperskov}, S.~A., {Moiseev}, A.~V., {Khoperskov}, A.~V., \& {Saburova},
  A.~S. 2014, \mnras, 441, 2650

\bibitem[{{Koribalski} {et~al.}(2020){Koribalski}, {Staveley-Smith},
  {Westmeier}, {Serra}, {Spekkens}, {Wong}, {Lee-Waddell}, {Lagos},
  {Obreschkow}, {Ryan-Weber}, {Zwaan}, {Kilborn}, {Bekiaris}, {Bekki},
  {Bigiel}, {Boselli}, {Bosma}, {Catinella}, {Chauhan}, {Cluver}, {Colless},
  {Courtois}, {Crain}, {de Blok}, {D{\'e}nes}, {Duffy}, {Elagali}, {Fluke},
  {For}, {Heald}, {Henning}, {Hess}, {Holwerda}, {Howlett}, {Jarrett}, {Jones},
  {Jones}, {J{\'o}zsa}, {Jurek}, {J{\"u}tte}, {Kamphuis}, {Karachentsev},
  {Kerp}, {Kleiner}, {Kraan-Korteweg}, {L{\'o}pez-S{\'a}nchez}, {Madrid},
  {Meyer}, {Mould}, {Murugeshan}, {Norris}, {Oh}, {Oosterloo}, {Popping},
  {Putman}, {Reynolds}, {Rhee}, {Robotham}, {Ryder}, {Schr{\"o}der}, {Shao},
  {Stevens}, {Taylor}, {van{\^A} der Hulst}, {Verdes-Montenegro}, {Wakker},
  {Wang}, {Whiting}, {Winkel}, \& {Wolf}}]{Koribalski2020}
{Koribalski}, B.~S., {Staveley-Smith}, L., {Westmeier}, T., {et~al.} 2020,
  \apss, 365, 118

\bibitem[{{Krajnovi{\'c}} {et~al.}(2006){Krajnovi{\'c}}, {Cappellari}, {de
  Zeeuw}, \& {Copin}}]{Krajnovic2005}
{Krajnovi{\'c}}, D., {Cappellari}, M., {de Zeeuw}, P.~T., \& {Copin}, Y. 2006,
  \mnras, 366, 787

\bibitem[{{Lagos} {et~al.}(2018){Lagos}, {Tobar}, {Robotham}, {Obreschkow},
  {Mitchell}, {Power}, \& {Elahi}}]{Lagos2018}
{Lagos}, C. d.~P., {Tobar}, R.~J., {Robotham}, A. S.~G., {et~al.} 2018, \mnras,
  481, 3573

\bibitem[{{Lelli} {et~al.}(2016){Lelli}, {McGaugh}, \& {Schombert}}]{Lelli2016}
{Lelli}, F., {McGaugh}, S.~S., \& {Schombert}, J.~M. 2016, \aj, 152, 157

\bibitem[{{Lelli} {et~al.}(2019){Lelli}, {McGaugh}, {Schombert}, {Desmond}, \&
  {Katz}}]{Lelli2019}
{Lelli}, F., {McGaugh}, S.~S., {Schombert}, J.~M., {Desmond}, H., \& {Katz}, H.
  2019, \mnras, 484, 3267

\bibitem[{{Lewis}(2019)}]{Lewis2019}
{Lewis}, C. 2019, PhD thesis, Queen's University at Kingston, Canada

\bibitem[{{McGaugh} {et~al.}(2000){McGaugh}, {Schombert}, {Bothun}, \& {de
  Blok}}]{McGaugh2000}
{McGaugh}, S.~S., {Schombert}, J.~M., {Bothun}, G.~D., \& {de Blok}, W.~J.~G.
  2000, \apjl, 533, L99

\bibitem[{{Mutabazi}(2021)}]{Mutabazi2021}
{Mutabazi}, T. 2021, \apj, 911, 16

\bibitem[{Nelder \& Mead(1965)}]{Nelder65}
Nelder, J.~A., \& Mead, R. 1965, Computer Journal, 7, 308

\bibitem[{{Oh} {et~al.}(2011){Oh}, {de Blok}, {Brinks}, {Walter}, \&
  {Kennicutt}}]{Oh2011}
{Oh}, S.-H., {de Blok}, W.~J.~G., {Brinks}, E., {Walter}, F., \& {Kennicutt},
  Robert~C., J. 2011, \aj, 141, 193

\bibitem[{{Oh} {et~al.}(2018){Oh}, {Staveley-Smith}, {Spekkens}, {Kamphuis}, \&
  {Koribalski}}]{Oh2018}
{Oh}, S.-H., {Staveley-Smith}, L., {Spekkens}, K., {Kamphuis}, P., \&
  {Koribalski}, B.~S. 2018, \mnras, 473, 3256

\bibitem[{{Papastergis} \& {Shankar}(2016)}]{Papastergis2016}
{Papastergis}, E., \& {Shankar}, F. 2016, \aap, 591, A58

\bibitem[{{Ponomareva} {et~al.}(2016){Ponomareva}, {Verheijen}, \&
  {Bosma}}]{Ponomareva2016}
{Ponomareva}, A.~A., {Verheijen}, M. A.~W., \& {Bosma}, A. 2016, \mnras, 463,
  4052

\bibitem[{Press {et~al.}(1992)Press, Teukolsky, Vetterling, \&
  Flannery}]{Press1992}
Press, W.~H., Teukolsky, S.~A., Vetterling, W.~T., \& Flannery, B.~P. 1992,
  Numerical Recipes in C, 2nd edn. (Cambridge, USA: Cambridge University Press)

\bibitem[{Reback {et~al.}(2020)Reback, McKinney, jbrockmendel, den Bossche,
  Augspurger, Cloud, gfyoung, Sinhrks, Klein, Roeschke, Hawkins, Tratner, She,
  Ayd, Petersen, Garcia, Schendel, Hayden, MomIsBestFriend, Jancauskas,
  Battiston, Seabold, chris b1, h~vetinari, Hoyer, Overmeire, alimcmaster1,
  Dong, Whelan, \& Mehyar}]{PANDAS}
Reback, J., McKinney, W., jbrockmendel, {et~al.} 2020, pandas-dev/pandas:
  Pandas 1.0.3, doi:\url{10.5281/zenodo.3715232}

\bibitem[{{Reynolds} {et~al.}(2021){Reynolds}, {Westmeier}, {Elagali},
  {Catinella}, {Cortese}, {Deg}, {For}, {Kamphuis}, {Kleiner}, {Koribalski},
  {Lee-Waddell}, {Oh}, {Rhee}, {Serra}, {Spekkens}, {Staveley-Smith},
  {Stevens}, {Taylor}, {Wang}, \& {Wong}}]{Reynolds2021}
{Reynolds}, T.~N., {Westmeier}, T., {Elagali}, A., {et~al.} 2021, \mnras, 505,
  1891

\bibitem[{{Reynolds} {et~al.}(2022){Reynolds}, {Catinella}, {Cortese},
  {Westmeier}, {Meurer}, {Shao}, {Obreschkow}, {Rom{\'a}n},
  {Verdes-Montenegro}, {Deg}, {D{\'e}nes}, {For}, {Kleiner}, {Koribalski},
  {Lee-Waddell}, {Murugeshan}, {Oh}, {Rhee}, {Spekkens}, {Staveley-Smith},
  {Stevens}, {van der Hulst}, {Wang}, {Wong}, {Holwerda}, {Bosma}, {Madrid}, \&
  {Bekki}}]{Reynolds2022}
{Reynolds}, T.~N., {Catinella}, B., {Cortese}, L., {et~al.} 2022, \mnras, 510,
  1716

\bibitem[{{Rogstad} {et~al.}(1974){Rogstad}, {Lockhart}, \&
  {Wright}}]{Rogstad1974}
{Rogstad}, D.~H., {Lockhart}, I.~A., \& {Wright}, M.~C.~H. 1974, \apj, 193, 309

\bibitem[{{Sancisi} \& {Allen}(1979)}]{Sancisi1979}
{Sancisi}, R., \& {Allen}, R.~J. 1979, \aap, 74, 73

\bibitem[{{Schoenmakers} {et~al.}(1997){Schoenmakers}, {Franx}, \& {de
  Zeeuw}}]{Schoenmakers97}
{Schoenmakers}, R.~H.~M., {Franx}, M., \& {de Zeeuw}, P.~T. 1997, \mnras, 292,
  349

\bibitem[{{Sellwood} {et~al.}(2021){Sellwood}, {Spekkens}, \&
  {Eckel}}]{Sellwood2021}
{Sellwood}, J.~A., {Spekkens}, K., \& {Eckel}, C.~S. 2021, \mnras, 502, 3843

\bibitem[{{Serra} {et~al.}(2015){Serra}, {Westmeier}, {Giese}, {Jurek},
  {Fl{\"o}er}, {Popping}, {Winkel}, {van der Hulst}, {Meyer}, {Koribalski},
  {Staveley-Smith}, \& {Courtois}}]{Serra2015}
{Serra}, P., {Westmeier}, T., {Giese}, N., {et~al.} 2015, \mnras, 448, 1922

\bibitem[{{Spekkens} \& {Sellwood}(2007)}]{Spekkens2007}
{Spekkens}, K., \& {Sellwood}, J.~A. 2007, \apj, 664, 204

\bibitem[{{Starkman} {et~al.}(2018){Starkman}, {Lelli}, {McGaugh}, \&
  {Schombert}}]{Starkman2018}
{Starkman}, N., {Lelli}, F., {McGaugh}, S., \& {Schombert}, J. 2018, \mnras,
  480, 2292

\bibitem[{{Stevens} {et~al.}(2016){Stevens}, {Croton}, \&
  {Mutch}}]{Stevens2016}
{Stevens}, A. R.~H., {Croton}, D.~J., \& {Mutch}, S.~J. 2016, \mnras, 461, 859

\bibitem[{{Swaters} {et~al.}(2002){Swaters}, {van Albada}, {van der Hulst}, \&
  {Sancisi}}]{Swaters2002}
{Swaters}, R.~A., {van Albada}, T.~S., {van der Hulst}, J.~M., \& {Sancisi}, R.
  2002, \aap, 390, 829

\bibitem[{{Tamburro} {et~al.}(2009){Tamburro}, {Rix}, {Leroy}, {Mac Low},
  {Walter}, {Kennicutt}, {Brinks}, \& {de Blok}}]{Tamburro2009}
{Tamburro}, D., {Rix}, H.~W., {Leroy}, A.~K., {et~al.} 2009, \aj, 137, 4424

\bibitem[{{Tully} \& {Fisher}(1977)}]{Tully1977}
{Tully}, R.~B., \& {Fisher}, J.~R. 1977, \aap, 54, 661

\bibitem[{{Tully} {et~al.}(2013){Tully}, {Courtois}, {Dolphin}, {Fisher},
  {H{\'e}raudeau}, {Jacobs}, {Karachentsev}, {Makarov}, {Makarova},
  {Mitronova}, {Rizzi}, {Shaya}, {Sorce}, \& {Wu}}]{tully2013}
{Tully}, R.~B., {Courtois}, H.~M., {Dolphin}, A.~E., {et~al.} 2013, \aj, 146,
  86

\bibitem[{{van Albada} {et~al.}(1985){van Albada}, {Bahcall}, {Begeman}, \&
  {Sancisi}}]{vanAlbada1985}
{van Albada}, T.~S., {Bahcall}, J.~N., {Begeman}, K., \& {Sancisi}, R. 1985,
  \apj, 295, 305

\bibitem[{{van Albada} \& {Sancisi}(1986)}]{vanAlbada1986}
{van Albada}, T.~S., \& {Sancisi}, R. 1986, Philosophical Transactions of the
  Royal Society of London Series A, 320, 447

\bibitem[{{van der Hulst} {et~al.}(1992){van der Hulst}, {Terlouw}, {Begeman},
  {Zwitser}, \& {Roelfsema}}]{vanderHulst1992}
{van der Hulst}, J.~M., {Terlouw}, J.~P., {Begeman}, K.~G., {Zwitser}, W., \&
  {Roelfsema}, P.~R. 1992, in Astronomical Society of the Pacific Conference
  Series, Vol.~25, Astronomical Data Analysis Software and Systems I, ed. D.~M.
  {Worrall}, C.~{Biemesderfer}, \& J.~{Barnes}, 131

\bibitem[{{Verheijen}(2001)}]{Verheijen2001}
{Verheijen}, M. A.~W. 2001, \apj, 563, 694

\bibitem[{Virtanen {et~al.}(2020)Virtanen, Gommers, Oliphant, Haberland, Reddy,
  Cournapeau, Burovski, Peterson, Weckesser, Bright, {van der Walt}, Brett,
  Wilson, Millman, Mayorov, Nelson, Jones, Kern, Larson, Carey, Polat, Feng,
  Moore, {VanderPlas}, Laxalde, Perktold, Cimrman, Henriksen, Quintero, Harris,
  Archibald, Ribeiro, Pedregosa, {van Mulbregt}, \& {SciPy 1.0
  Contributors}}]{SciPy}
Virtanen, P., Gommers, R., Oliphant, T.~E., {et~al.} 2020, Nature Methods, 17,
  261

\bibitem[{{Walter} {et~al.}(2008){Walter}, {Brinks}, {de Blok}, {Bigiel},
  {Kennicutt}, {Thornley}, \& {Leroy}}]{Walter2008}
{Walter}, F., {Brinks}, E., {de Blok}, W.~J.~G., {et~al.} 2008, \aj, 136, 2563

\bibitem[{{Wang} {et~al.}(2016){Wang}, {Koribalski}, {Serra}, {van der Hulst},
  {Roychowdhury}, {Kamphuis}, \& {Chengalur}}]{Wang16}
{Wang}, J., {Koribalski}, B.~S., {Serra}, P., {et~al.} 2016, \mnras, 460, 2143

\bibitem[{{Wang} {et~al.}(2021){Wang}, {Staveley-Smith}, {Westmeier},
  {Catinella}, {Shao}, {Reynolds}, {For}, {Lee}, {Liang}, {Wang}, {Elagali},
  {D{\'e}nes}, {Kleiner}, {Koribalski}, {Lee-Waddell}, {Oh}, {Rhee}, {Serra},
  {Spekkens}, {Wong}, {Bekki}, {Bigiel}, {Courtois}, {Hess}, {Holwerda},
  {McQuinn}, {Pandey-Pommier}, {van der Hulst}, \&
  {Verdes-Montenegro}}]{Wang2021}
{Wang}, J., {Staveley-Smith}, L., {Westmeier}, T., {et~al.} 2021, \apj, 915, 70

\bibitem[{{Westmeier} {et~al.}(2021){Westmeier}, {Kitaeff}, {Pallot}, {Serra},
  {van der Hulst}, {Jurek}, {Elagali}, {For}, {Kleiner}, {Koribalski},
  {Lee-Waddell}, {Mould}, {Reynolds}, {Rhee}, \&
  {Staveley-Smith}}]{Westmeier2021}
{Westmeier}, T., {Kitaeff}, S., {Pallot}, D., {et~al.} 2021, \mnras, 506, 3962

\bibitem[{{Westmeier} {et~al.}(2022){Westmeier}, {Deg}, {Spekkens}, {Reynolds},
  {Shen}, {Gaudet}, {Goliath}, {Huynh}, {Venkataraman}, {Lin}, {O’Beirne},
  {Catinella}, {Cortese}, {D\'{e}nes}, {Elagali}, {For}, {J\'{o}zsa},
  {Howlett}, M., J., {Kamphuis}, {Kilborn}, {Kleiner}, {Koribalski},
  {Lee-Waddell}, {Murugeshan}, {Rhee}, {Serra}, {Shao}, {Staveley-Smith},
  {Wang}, {Wong}, {Zwaan}, {Allison}, {Anderson}, {Ball}, {Bock}, {Brodrick},
  {Bunton}, {Cooray}, {Gupta}, {Hayman}, {Mahony}, {Moss}, {Ng}, {Pearce},
  {Raja}, {Roxby}, {Voronkov}, {Warhurst}, {Courtois}, \&
  {Said}}]{SoFiARelease}
{Westmeier}, T., {Deg}, N., {Spekkens}, K., {et~al.} 2022, \pasa,
  doi:\url{doi.org/10.1017/pasa.2022.50}

\bibitem[{{Whiting}(2012)}]{Whiting2012}
{Whiting}, M.~T. 2012, \mnras, 421, 3242

\end{thebibliography}

\end{document}